%% file: ms.tex
\documentclass[oupdraft]{bio}
\pdfoutput=1
\usepackage{color}

\makeatletter
\DeclareRobustCommand{\rvdots}{%
  \vbox{
    \baselineskip4\p@\lineskiplimit\z@
    \kern-\p@
    \hbox{.}\hbox{.}\hbox{.}
  }}
\makeatother

\input{abbreviations}

\begin{document}

\title{Correlation-Adjusted Regression Survival Scores for High-Dimensional Variable Selection}

\author{Thomas Welchowski*$^1$, Verena Zuber$^2$, Matthias Schmid$^1$ \\[4pt]
\textit{$^1$ Department of Medical Biometry, Informatics and Epidemiology, \\ University Hospital Bonn, Germany}
\\[2pt]
\textit{$^2$ MRC Biostatistics Unit, Cambridge University, United Kingdom}
\\[2pt]
{welchow@imbie.meb.uni-bonn.de, verenaz@mrc-bsu.cam.ac.uk, matthias.schmid@imbie.uni-bonn.de}}

\markboth%
{Welchowski, T. and Zuber, V. and Schmid, M.}
{High dimensional CARS}

\maketitle

\footnotetext{To whom correspondence should be addressed.}

\begin{abstract}
{\noindent {\it Background} \\
\noindent The development of classification methods for personalized medicine is highly dependent on the identification of predictive genetic markers. In survival analysis it is often necessary to discriminate between influential and non-influential markers. Usually, the first step is to perform a univariate screening step that ranks the markers according to their associations with the outcome. It is common to perform screening using Cox scores, which quantify the associations between survival  and each of the markers individually. Since Cox scores do not account for dependencies between the markers, their use is suboptimal in the presence highly correlated markers. \\[0.15cm]
\noindent {\it Methods} \\
\noindent As an alternative to the Cox score, we propose the correlation-adjusted regression survival (CARS) score for right-censored survival outcomes. By removing the correlations between the markers, the CARS score quantifies the associations between the outcome and the set of ``de-correlated'' marker values. Estimation of the scores is based on inverse probability weighting, which is applied to log-transformed event times. For high-dimensional data, estimation is based on shrinkage techniques. \\[0.15cm]
\noindent {\it Results} \\
\noindent The consistency of the CARS score is proven under mild regularity conditions. In simulations, survival models based on CARS score rankings achieved higher areas under the precision-recall curve than competing methods. Two example applications on prostate and breast cancer confirmed these results. CARS scores are implemented in the \textit{R} package \textit{carSurv}. \\[0.15cm]
\noindent {\it Conclusions} \\
\noindent In research applications involving high-dimensional genetic data, the use of CARS scores for marker selection is a favorable alternative to Cox scores even when correlations between covariates are low. Having a straightforward interpretation and low computational requirements, CARS scores are an easy-to-use screening tool in personalized medicine research. }
{Biomarker discovery; Breast cancer; Multi-gene signature; Personalized medicine; Prostate cancer; Survival modeling}
\end{abstract}

\input{1-Introduction}
\input{2-Methods}
\input{3-Results}
\input{4-End}

\bibliographystyle{biorefs}
\bibliography{survival}

\end{document}

%% file: abbreviations.tex
\newcommand\bI{\boldsymbol I}

\newcommand\bmu{\boldsymbol \mu}

\newcommand\bP{\boldsymbol P}
\newcommand\bR{\boldsymbol R}

\newcommand\bSigma{\boldsymbol \Sigma}

\newcommand\bX{\boldsymbol X}
\newcommand\bx{\boldsymbol x}

\newcommand\bbeta{\boldsymbol \beta}

\newcommand\btheta{\boldsymbol \theta}

\newcommand\Var{\text{Var}}


%% file: 1-Introduction.tex
\section{\textbf {Introduction}}
\label{sec1}

One of the key issues in personalized medicine is to identify genetic marker signatures for the planning and the prognosis of targeted cancer therapies. With more than one of three people developing some form of cancer during their lifetimes \citep{CancerRates}, individualized therapies based on genetic markers are expected to play a major role in improving progression-free and overall survival of cancer patients. Among men, for example, prostate cancer is the cancer with the highest prevalance. While there are several clinical models available for predicting disease progression, it remains a challenging task to develop molecular signatures and improve predictive accuracy of existing models \citep{Sboner2010}. \\

Since cancer research is heavily focused on time-to-event outcomes such as progression-free survival, metastasis-free survival and/or overall survival, survival analysis is one of the predominant statistical approaches to analyze data collected in clinical cancer trials. When the aim is to relate a time-to-event outcome to a set of predictors (e.g., clinical information or genetic markers), it is common to use a survival model such as the proportional hazards model by Cox. However, when data are high-dimensional (e.g., when the number of measured genetic markers exceeds the sample size), it is impossible to fit a Cox regression model including all available covariates. A solution to this problem could be to use regularized methods such as ridge-penalized Cox regression, but even these methods often break down when the number of available markers (in particular, the number of non-influential markers) is large. Is therefore common practice to carry out data-driven variable selection {\em before} fitting the survival model, and to include only those ``influential markers'' that have passed selection step. \\

The predominant method for variable selection in cancer research is {\em univariate screening}, which evaluates the associations between the outcome of a trial and each covariate separately, e.g., by computing correlation coefficients or fitting simple regression models. The coefficients of association are usually ranked by magnitude, and the most highly ranked covariates are selected for inclusion in the statistical model of interest. \cite{sureIndScreen} have provided a theoretical justification for this approach by showing that univariate screening is suitable to identify influential covariates with high probabilities under mild regularity conditions. Still, a major problem of this approach is that associations between covariates are ignored, and that the set of selected markers may include many redundant markers if these are highly correlated (and hence carry the same information regarding their associations with the outcome). While decreasing the robustness of the final statistical model, a large set of redundant markers will also cause influential markers with weaker univariate associations to be dropped from the model. This information loss is particularly problematic when the number of selected markers needs to be restricted to a small value due to sample size or cost limitations. \\

In survival analysis with a right-censored time-to-event outcome, univariate screening is mostly done by computing {\em Cox scores}, which are given by either the Z scores obtained from univariate Cox regression models or by the p-values obtained from the respective likelihood ratio or score tests. Although p-values can be corrected for multiple testing (e.g. see \cite{FDR}) to identify informative covariates, Cox scores share the same disadvantages as the univariate screening methods mentioned above. \\

To address these problems, we consider the {\em correlation-adjusted regression (CAR) score} approach, which provides a criterion for variable ranking that is based on the de-correlation of covariates in linear regression. By applying a Mahalanobis-type ``de-correlating'' transformation to the covariates, CAR scores measure the correlations between the de-correlated variables and the continuous outcome. The set of correlation coefficients defines a ranking of the covariates, which can be used to select informative variables in the same way as with the univariate screening methods described above. As the correlation coefficients among the covariates tend to zero, CAR scores become identical to the correlations between the non-transformed covariates and the outcome. In simulations for linear regression, CAR scores were shown to outperform methods for regularized regression (boosting, lasso) with regard to their ability to correctly recover causal genetic markers and their rankings \citep{ZuberSNP2012}. On the other hand, CAR scores have not been used to model time-to-event outcomes in cancer research, as an extension of the CAR approach to right-censored data has been lacking so far. \\

The purpose of this paper is therefore to develop a CAR-based method for ranking high-dimensional sets of genetic marker variables in survival analysis. The resulting score, which in the following will be termed {\em correlation-adjusted regression survival (CARS) score}, quantifies the correlations between the log-transformed survival time $Y=\log(T)$ and the de-correlated set of covariates $\boldsymbol{X}$. We will first define a theoretical version of the CARS score on the population level (Section \ref{popLevel}). Afterwards, we will provide details on the estimation of the scores from a sample of right-censored data (Section \ref{sampleLevel}). Specifically, we will show that all relevant expressions can be estimated using inverse-probability-of-censoring (IPC) weighting techniques, as proposed by \cite{van2012unified}. In Section \ref{sec3}, we will present the results of a simulation study that was carried out to compare the CARS approach to univariate screening based on Cox scores. In addition, we will apply CARS scores to the Swedish Watchful Waiting Cohort data \citep{Sboner2010} and to a data set on invasive breast cancer \citep{doi:10.1001/jama.2011.593}. The results of the paper will be summarized in Section \ref{sec4}.

%% file: 2-Methods.tex
\section{\textbf {Methods}}
\label{sec2}


\subsection{\textbf {Full data world / population level}} \label{popLevel}

The main focus of survival modeling is on analyzing the effects of a set of covariates ${\boldsymbol x} \in \mathbb{R}^d$ on a survival time $T \in \mathbb{R}^+$. We assume that 
the vector ${\boldsymbol x} = (X_1, \dots , X_d)^\top$ has expectation $\bmu$, covariance matrix $\bSigma$ and correlation matrix $\bP_{\bX}$. Similarly, we assume that the survival time $T$ has a finite expectation $\mu_T$ and variance $\sigma^2_T$. A popular approach to quantify the relationship between $T$ and ${\boldsymbol x}$ is the parametric accelerated failure time (AFT) model \citep{klein2013handbook}, which is based on log-transformed survival time $Y :=\log(T)$ and the model equation
\begin{equation}\label{log-lm}
Y = \beta_0 + \boldsymbol {x}^T \bbeta + \epsilon \, ,
\end{equation}
where $\bbeta \in \mathbb{R}^d$ is a vector of regression coefficients and $\epsilon$ a noise variable. For the derivation of CARS scores we will consider the special case of log-normally distributed survival times, i.e.\@ $\epsilon$ is assumed to follow a normal distribution with zero mean and constant variance. Then the expected squared prediction error $\mbox{E}\left( ( Y - \beta_0 - \boldsymbol {x}^T \bbeta )^2\right)$ is minimized by regression coefficients equal to
\begin{equation} \label{eq:bcoef}
\bbeta^*  = \Sigma^{-1}\Sigma_{\bX Y}  \, , 
\end{equation} 
where $\Sigma_{\bX Y}$ is the $d$-dimensional vector of
covariances between ${\boldsymbol x}$ and $Y$, and the intercept
\begin{equation}
\label{eq:acoef}
\beta_0 ^* = \mu_Y - \bbeta^T \bmu \,. 
\end{equation}
Model \eqref{log-lm} is a Gaussian linear regression model. Therefore, in the absence of censoring, a measure of variable importance is the CAR score $\btheta$ \citep{highDimVarCAR} defined by
\begin{equation}\label{omega-car}
\btheta = \bP_{\bX}^{-1/2} \bP_{\bX, Y} \, ,
\end{equation}
where $\bP_{\bX} \in \mathbb{R}^{d\times d}$ is the correlation matrix of the covariates $\bX$ and $\bP_{\bX, Y}  \in \mathbb{R}^{d\times 1}$ is the vector of correlations between 
the covariates $\bX$ and the log-transformed survival time $Y=\log(T)$. Analogous to Cox scores, the components of $\btheta$ can be ordered by magnitude to give an importance ranking of the covariates. \\

The original CAR score for Gaussian linear regression can be interpreted as the correlations between the outcome variable and the de-correlated covariates, which are defined by the orthogonal transformation $\boldsymbol{z} = \boldsymbol{P}_{\bX}^{-1/2} \boldsymbol{x}$ \citep{highDimVarCAR}. Using the best linear unbiased predictor $Y^\star = \beta_0^* + \boldsymbol{x}^T {\boldsymbol{\beta}}^*$ derived from Estimators \eqref{eq:bcoef} and \eqref{eq:acoef}, and defining $\sigma^2_Y := \mbox{Var}(Y)$, the total variance of $Y$ can be decomposed as follows:
\begin{align} \label{new1}
\overbrace{\Var(Y)}^{\text{Total variance}}  = &   \overbrace{\Var(Y^\star)}^{\text{Explained variance}}  + \overbrace{\Var(Y-Y^\star)}^{\text{Unexplained variance}} \\ \label{new2}
= & \quad \sigma_Y^2 \boldsymbol{P}_{Y, Y^*}^2 \quad + 
\quad\sigma_Y^2 (1-\boldsymbol{P}_{Y, Y^*}^2) \, ,
\end{align}
where
\begin{equation}
\boldsymbol{P}_{Y, Y^*}^2 = \boldsymbol{P}_{\bX , Y}^\top \boldsymbol{P}_{\bX}^{-1} \boldsymbol{P}_{\bX , Y} \label{total-var-decomp} = \btheta^T  \btheta
\end{equation}
is the squared correlation between $Y$ and $Y^*$. From Equations \eqref{new1} to \eqref{total-var-decomp} it follows that the CAR score is the central quantity to assess which variables contribute to the explained variance, or equivalently reduce the unexplained variance. Importantly, $ \boldsymbol{P}_{ \bX}^{-1/2} \boldsymbol{x}$ is not the Mahalanobis transform but another form of de-correlation that has the advantageous feature of maximizing the correlation between the de-correlated covariates and the standardized original covariates \citep{optWhiteDec}. In contrast, the Mahalanobis transform maximizes the cross-covariance between the de-correlated covariates and the original covariates. \cite{highDimVarCAR} and \cite{ZuberSNP2012} demonstrated that estimated CAR scores result in improved variable rankings and a high predictive performance when compared to other variable selection and modeling techniques such as lasso or boosting. Although these estimated scores work well for regression models with a continuous outcome, they were not able to deal with right-censored data so far. We will therefore develop the CAR survival (CARS) score that extends traditional estimators of CAR scores to survival modeling.

\subsection{\textbf {Observed data world / sample level}} \label{sampleLevel}

In the observed data world one often has to deal with right-censoring, i.e.\@ one is no longer able to observe the uncensored survival times of all observations but only the minimum of the true survival time $T$ and a censoring time $C \in\mathbb{R}^+$. The observed variable of interest is then defined by $\tilde{Y}=\log(\tilde{T})$, where $\tilde{T} :=\min(T,C)$. Additionally, we introduce a status indicator $\Delta := \mathrm{I} (T \leq C)$, i.e., $\Delta=1$ if the event is observed and $\Delta=0$ if censoring has occurred. We will further assume that $T$ and $C$ are independent random variables. At the sample level the empirical CAR score is defined by
\begin{equation}\label{omega-car2}
\hat{\btheta} = \bR_{\text{shrink}}^{-1/2} \, \bR_{\bX, Y}  \, , 
\end{equation}
where $\bR_{\text{shrink}} \in \mathbb{R}^{d\times d}$ is a shrinkage estimator of the correlation matrix $\bP_{\bX}$ and $\bR_{\bX, Y} \in \mathbb{R}^{d\times 1}$ is an estimator of the vector of correlations $\bP_{\bX, Y}$. The definition of $\bR_{\text{shrink}}$ will be provided below. In contrast to uncensored Gaussian regression, standard estimation of $\bP_{\bX, Y}$ using the sample correlations between $\bx$ and $\tilde{Y}$ is no longer appropriate, as this would result in biased estimators in the presence of right-censoring. To overcome this problem, we suggest to adjust the sample correlations by inverse-probability-of-censoring (IPC) weighting 
\citep{van2012unified}, which will result in a consistent estimator of $\bP_{\bX, Y}$. \\

 {\it Definition of IPC weigths for right-censored data}:
 Let $\tilde{T}_1 , \ldots , \tilde{T}_n $ be the observed values of $\tilde{T}$ and ${C}_1 , \ldots , {C}_n $ the underlying censoring times in a sample of i.i.d.\@ observations of size $n$. Following \cite{van2012unified}, we define the IPC weight of the $i$-th observation by 
\begin{equation}\label{ipcweightsdef}
w_{i} := \frac{\Delta_i }{\hat{G}_n (\log( \tilde{T}_i))} \ , \ \  i\in \{ 1,\dots, n \}\, ,
\end{equation}
where $\Delta_i$, $i=1,\ldots , n$, are the sample values of $\Delta$ and $\hat{G}_n (\log( \tilde{T}_i))$ is an estimate of the survivor function $G(\log(  T_i))$ of the logarithmic censoring process, i.e.\@ the probability
\begin{equation}\label{survivor}
G(\log (\tilde{T}_i)) = P(\log (C_i) > \log(\tilde{T}_i)) \, .
\end{equation}
By definition, censored observations ($\Delta_i=0$) result in zero IPC weights. In line with \cite{van2012unified}, we further assume that $G(\cdot ) > \nu > 0$, where $\nu$ is a small positive real number (this assumption will become important in the consistency proof in Appendix \ref{proofWeightedMean}). To compute $\hat{G}_n(\cdot )$, we apply the Kaplan-Meier estimator to the observed logarithmic survival times $\tilde{T}_1,\ldots , \tilde{T}_n$, using the event indicators $1 - \Delta_i$, $i=1,\ldots , n$. \\

 {\em Estimation of the correlation vector $\bP_{\bX, Y}$ using IPC weighting}: 
The estimation of $\bP_{\bX, Y}$ comprises of the following steps:
\begin{enumerate}
\item Estimate the expectations of the covariates $X_1, \ldots , X_d$ by their empirical means $\bar{X}_j =\sum_{i=1}^n X_{ij} / n$, $j = 1, \ldots, d$, respectively, where $X_{ij}$ denotes the sample value of the $j$-th covariate in observation $i$. Similarly, estimate the variances of $X_1, \ldots , X_d$ by
their sample variances $S_j^2 = \sum_{i=1}^n \left( X_{ij} - \bar{X}_j \right)^2 /(n-1)$, $j = 1, \ldots , d$.
\item Estimate the expectation of $Y$ by the weighted mean
\begin{equation}
\label{expY}
\bar{Y}_w = \frac{1}{n} \sum_{i=1}^n w_{i} \log (\tilde{T}_{i} ) \ ,
\end{equation}
where $w_i$ , $i=1,\ldots, n$ are the IPC weights defined in Equation \eqref{ipcweightsdef}.
Similarly, estimate the variance of $Y$ by
\begin{equation}
\label{varY}
S_{Y;w}^2 = \frac{1}{n} \sum_{i=1}^n w_{i} \left( \log (\tilde{T} _{i} ) - \bar{Y}_w \right)^2 \ . 
\end{equation}
\item The covariance of $X_j$ and $Y$ is again estimated by IPC weighting:
\begin{equation}
\label{covXY}
S_{X_j,Y;w} = \frac{1}{n} \sum_{i=1}^n w_{i} (X_{ij} - \bar{X}_j) \left( \log (\tilde{T}_{i} ) - \bar{Y}_w \right) \ , \ \ j= 1,\ldots , d. 
\end{equation}
\item The final step is to compute the empirical correlation vector $\bR_{\bX, Y}$ by combining the estimators defined in Steps 1 to 3 above:
\begin{equation}
\label{Phat}\bR_{\bX, Y} = \left(
\frac{S_{X_j,Y;w}}
{\sqrt{S_{X_j}^2} \sqrt{S_{Y;w}^2}}
\right)_{j = 1, \dots , d} \ .
\end{equation}
\end{enumerate}

\vspace{.5cm}
 {\em Estimation of the correlation matrix $\bP_{\bX}^{-1/2}$}: Since the data values of the covariates are not affected by censoring, the usual sample variance-covariance estimators could be applied to obtain an estimate of $\bP_{\bX}^{-1/2}$. In the presence of high-dimensional data, however, these estimators usually break down. For example, the estimation of the $d \times d$ correlation matrix $\bP_{\bX}$ -- or more precisely its inverse square root -- is challenging when when $d$ is much larger than the sample size \citep{databaseNCBI, kolesnikov2014arrayexpress}. We therefore propose to employ a shrinkage  correlation estimator \citep{shrinkCovar, highDimVarCAR} to estimate $\bP_{\bX}$, which is given by
\begin{align}
& \bR_{\text{shrink}} = \lambda \bI_d + (1-\lambda) \bR_{\bX}  \, ,
\end{align}
where $\lambda \in \mathbb{R}^+$ is a shrinkage parameter, $\bI_d$ is the identity matrix of dimension $d\times d$, and $\bR_{\bX}$ is the matrix containing the empirical bivariate sample correlations of the covariates. Following \cite{shrinkCovar}, we define $\lambda := \sum_{j\ne k} \widehat{\mbox{Var}} (r_{j,k}) / \sum_{j\ne k} r_{j,k}^2$, where $r_{j,k}$ denotes the sample correlation between the $j$-th and the $k$-th covariate. The inverse square root $\bR_{\text{shrink}}^{-1/2}$ can be computed very efficiently by applying a singular value decomposition  of the sample correlation matrix. For details, we refer to \cite{highDimVarCAR} and \cite{optWhiteDec}. The estimation of $\bP_{\bX, Y}$ described above, combined with the shrinkage estimator $\bR_{\text{shrink}}$, defines the CARS score in Equation \eqref{omega-car2}. \\

 {\em Consistency of CARS scores}: Next we give a sketch of the consistency proof for the estimated CARS score $\hat{\btheta} = \bR_{\text{shrink}}^{-1/2} \, \bR_{\bX, Y}$. As shown in detail in Appendix \ref{proofCARS}, $\hat{\btheta}$ converges to its population value $\btheta = \bP_{\bX}^{-1/2} \, \bP_{\bX, Y}$ as $n\to\infty$, provided that (i) censoring is independent of the survival times, and (ii) the Kaplan-Meier estimator $\hat{G}_n$ is a consistent estimator of $G$. More specifically, by embedding the IPC-weighted expressions given in the Estimators \eqref{expY} to \eqref{Phat} in the framework of unbiased estimating equations \citep{huberEstEq, transWeightReg}, we show that each of the estimators contained in the definition of $\hat{\btheta}$ is a consistent estimator of its  respective population variance or covariance. As a consequence, $\hat{\btheta}$ results in a consistent estimator of the population value ${\btheta}$. \\

{\em Variable selection based on CARS scores}: As shown above, CARS scores measure the associations between the de-correlated covariates and the time-to-event outcome $T$. Variable selection can therefore be carried out by ranking the CARS scores according to their absolute values. A set of covariates is selected whose absolute CARS scores exceed the pre-defined threshold value $\phi$. A suitable threshold can, for example, be obtained by cross-validating the multivariable survival model that incorporates the selected covariates. In this paper, we will use a computationally less expensive strategy and apply the adaptive false discovery rate density approach proposed by \cite{fndrNullModel}, which assumes a two-component discrete mixture model of ``influential'' and ``non-influential'' covariates. Based on this model, a suitable threshold value $\phi$ can be estimated by a tradeoff between the false-non-discovery rate $P(\text{``influential''}|\theta \leq \phi)$ and the false-discovery rate $P(\text{``not-influential''}|\theta \geq \phi)$. The associated parameters of the mixture model are estimated by penalized maximum likelhood and a modified semi-parametric Grenander estimator \citep{grenanderDens, fndrNullModel}. For details, and also for an overview of the advantages of the approach, we refer to \cite{fndrNullModel}.

%% file: 3-Results.tex
\section{\textbf {Results}}
\label{sec3}

\subsection{\textbf {Design of the simulation study}} \label{simulation}

To analyze the performance of CARS scores, we compared the CARS-based screening approach to a univariate screening approach using Cox scores \citep{sureIndScreen}. With the latter approach, a univariate Cox model was fitted for each covariate, and the Cox scores were given by the standardized coefficients of these models. In addition, we fitted multivariable Cox regression models with $L_1$-penalized coefficients \citep{glmnetPub}. In these models, a variable was considered to be ``included'' in the model when its $L_1$-penalized coefficient estimate differed from zero. To ensure a fair comparison with the other methods, no tuning of the regularization parameter was applied, as this would have required additional test data. Instead we used the median value of the default $L_1$ norm regularization path computed by the \textit{R} \citep{Rsoftware} package \textit{glmnet} \citep{glmnetPub}. \\

We considered three sample sizes ($n \in \left\lbrace 250, 500, 1000 \right\rbrace$) and three dimensions of the covariate space ($d \in \left\lbrace 500, 1000, 2000 \right\rbrace$). The covariate values were generated from a multivariate normal distribution with zero mean. For the covariance structure of the multivariate normal distribution, a block correlation structure with three equally sized blocks was constructed. In the first block, $50\%$ of the correlations were set to $\rho = -0.25$, and the other $50\%$ were set to $\rho = 0.25$. In the second block, $50\%$ of the correlations were set to $\rho = 0.5$ and the other $50\%$ to $\rho = -0.5$, and in the third block, $50\%$ of the correlations were set to $\rho = 0.75$ and the other $50\%$ to $\rho = -0.75$. The correlation between covariates belonging to different blocks was zero. To satisfy the restrictions of a correlation matrix (e.g.\@ positive definiteness) the closest correlation matrix with regard to quadratic elementwise differences was computed. Further details on the algorithm for the construction of the correlation matrix are given in the Appendix \ref{constrCorr}. \\

The percentages of influential covariates that were related to the time-to-event outcome was varied according to the values $1\%$, $5\%$, and $10\%$. Two different correlation scenarios were analyzed: In the first scenario all influential variables were taken from the first block of the correlation matrix (``scenario with low absolute correlations''), and in the second scenario all influential variables were taken from the third block (``scenario with high absolute correlations''). The survival process was assumed to follow a log-normal distribution $T \sim \log N (\log\mu_T, \log\sigma_T^2)$ with expectation $\log\mu_T = \bX \bbeta$. The coefficients $\boldsymbol{\beta}$ were specified to be equidistant within the interval $\left[ -0.9, 1 \right]$, depending on the number of influential covariates. The variance $\log\sigma_T^2$ was adjusted such that explained variances of $\left\lbrace 25\%, 50\%, 75\% \right\rbrace$ were achieved on the log-scale. The censoring process was also assumed to be log-normally distributed. Its parameters were adjusted such that censoring rates of $0.25$ and $0.75$ were obtained. All values of the observed times that were higher than their empirical $90\%$ quantile were cut off and were set to be censored. For each of the scenarios we carried out $300$ independent simulation runs. \\ 

To evaluate the performance of the methods, the covariates selected by CARS scores, Cox scores and $L_1$ penalized Cox regression were compared to the sets of influential variables having a true (non-zero) effect on the survival times. For each threshold of the CARS and Cox scores, these comparisons resulted the cross-tabulation of the binary variables ``selected vs.\@ non-selected'' and ``influential vs.\@ non-influential''. In the case of $L_1$ penalized Cox regression variables with coefficients greater than zero were defined as selected and all other variables as non-selected. Since we were interested in detecting a small set of influential markers within a sparse modeling framework, we used the area under the precision-recall curve (PR-AUC, \citealt{informRetr}) to evaluate the cross tables and to measure the performance of the three methods. In addition, we investigated the ability of the methods to rank the variables - from least to most important - by analyzing the rank correlations between the true absolute coefficients and the corresponding estimated absolute CARS or Cox scores.

\subsection{\textbf {Scenario with low absolute correlations and low censoring rate} } \label{lowCorCase}

 \begin{figure}[!htbp]
 	\centering
   \includegraphics[width=\textwidth]{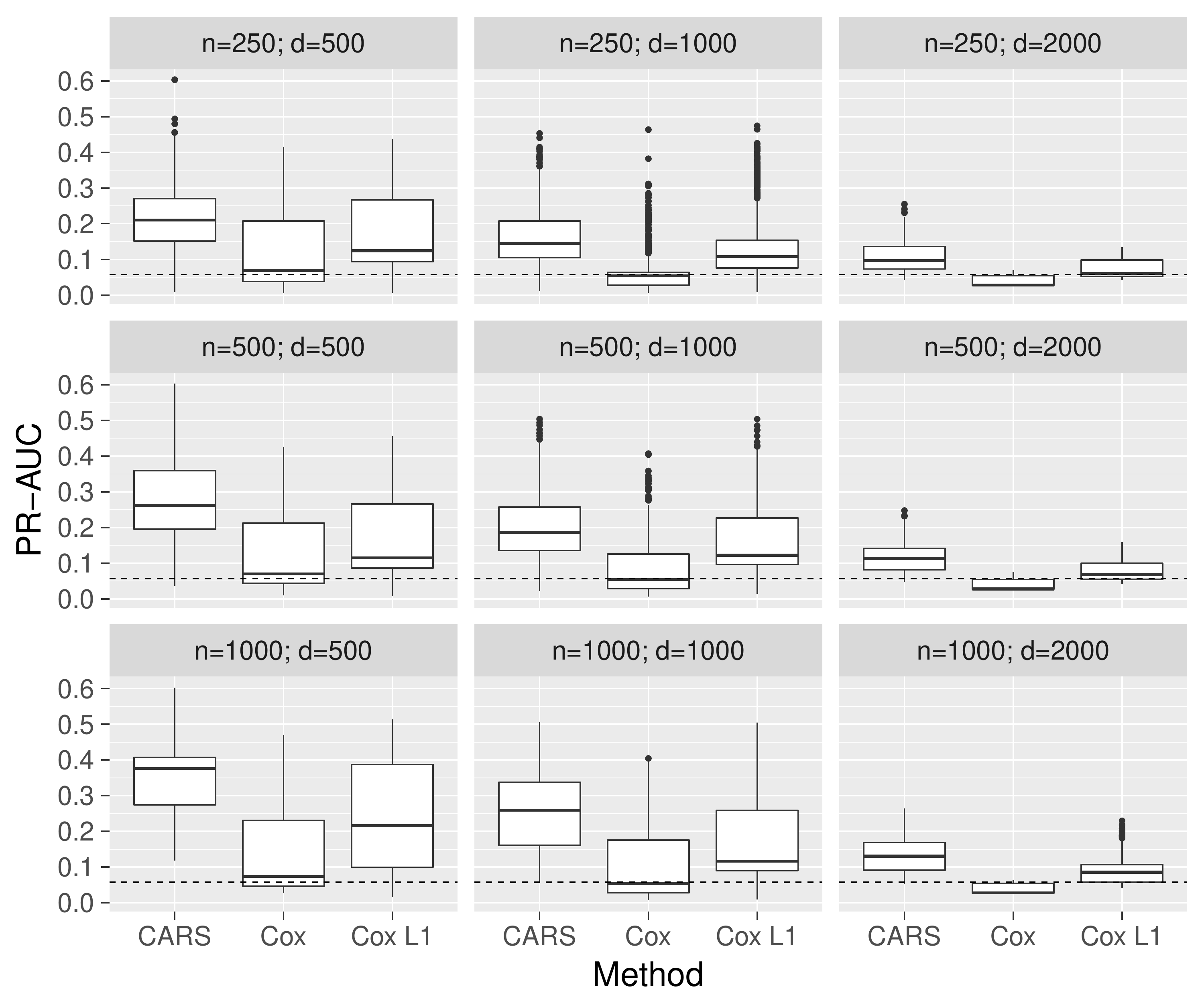}
 	\caption{Simulation: PR-AUC of CARS, Cox and Cox $L_1$ scores stratified by sample size (n) and number of covariates (d) with low absolute covariate correlations $(\rho=\pm 0.25)$ and low censoring rate of $25\%$. Each boxplot summarizes the results of $2700$ simulation runs (3 explained variance ratios x 3 signal to noise ratios x 300 repetitions).}
 	\label{simCARSlowCorPlotNxP}
 \end{figure}
 
  \begin{figure}[!htbp]
  	\centering
    \includegraphics[width=\textwidth]{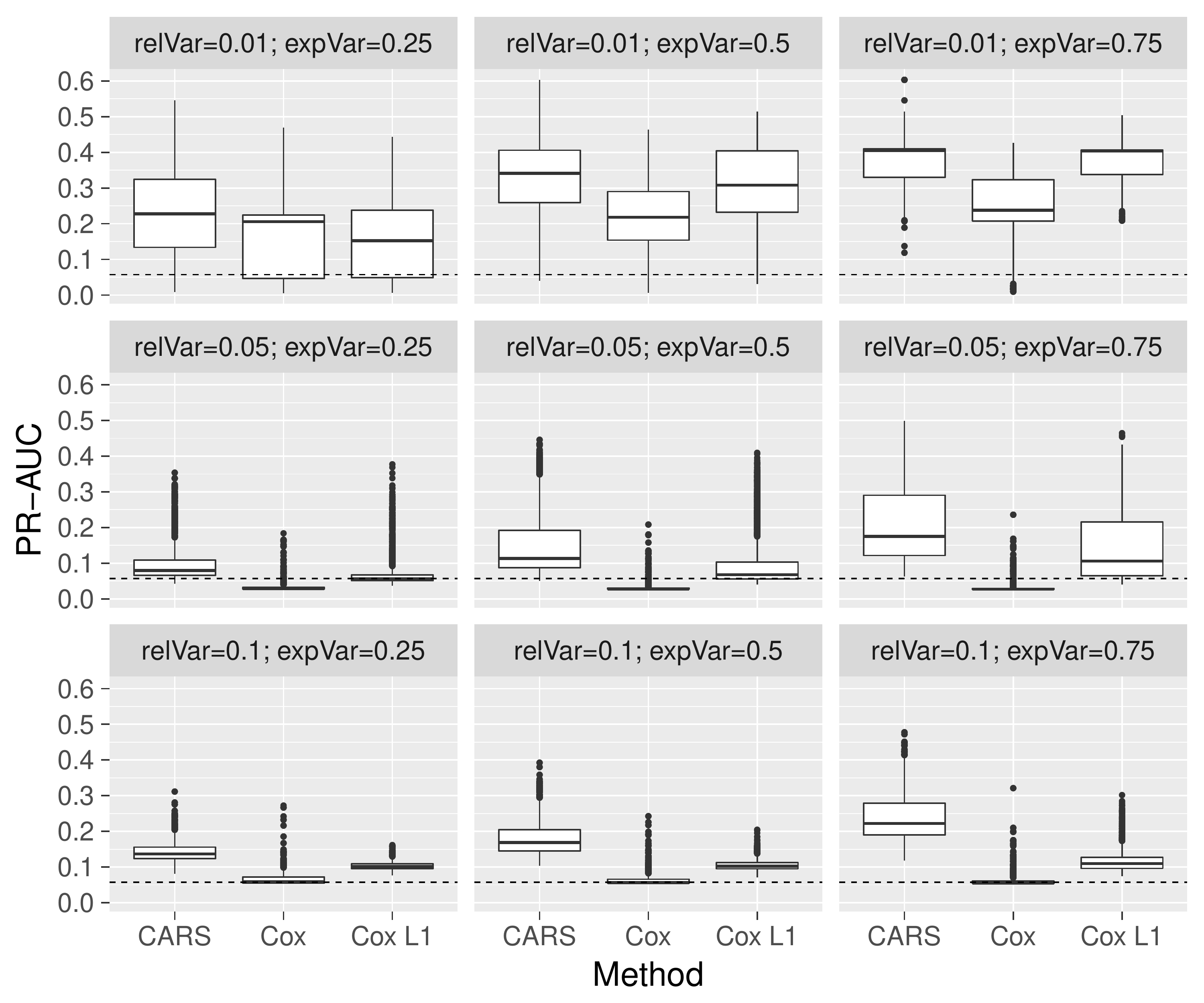}
  	\caption{Simulation: PR-AUC of CARS, Cox and Cox $L_1$ scores stratified by relative number of relevant variables (relVar) and explained variance (expVar) with low absolute covariate correlations $(\rho=\pm 0.25)$ and censoring rate of $25\%$. Each boxplot summarizes the results of $2700$ simulation runs (3 sample sizes x 3 number of covariates x 300 repetitions).}
  	\label{simCARSlowCorPlotNoVarExpVar}
  \end{figure}

The results of the scenario with low absolute correlations (first block of the correlation matrix, $\rho = \pm 0.25$) and with $25\%$ censoring rate are presented in this section. After running the algorithm for the construction of the correlation matrix (presented in Appendix \ref{constrCorr}), all correlations had absolute values that were smaller than $0.3$. Figure \ref{simCARSlowCorPlotNxP} shows the summary of all simulations results regarding sample size $n$ and number of covariates $d$. The median of CARS scores was higher than both Cox score approaches in most combinations of $n$ and $d$. In addition Figure \ref{simCARSlowCorPlotNoVarExpVar} displays the results with respect to the relative number of influential variables (relVar) and the explained variance (expVar). In most cases CARS scores had again higher median PR-AUC performance, except for the two upper right cases $\text{relVar}=0.01$ with $\text{expVar}=0.5$ or $\text{expVar}=0.75$. Higher signal to noise ratios increased the performance of CARS on average. The PR-AUC of CARS scores had the largest variability compared to the Cox approaches. $L_1$ penalized Cox scores showed larger variability of PR-AUC than Cox scores. Further CARS scores ranked influential variables better in the median than Cox scores, as shown in Figure \ref{figCor}. An overall summary is available in Figure \ref{figPRAUC} in the appendix.      

\begin{figure}
	\centering
  \includegraphics[scale = 0.3]{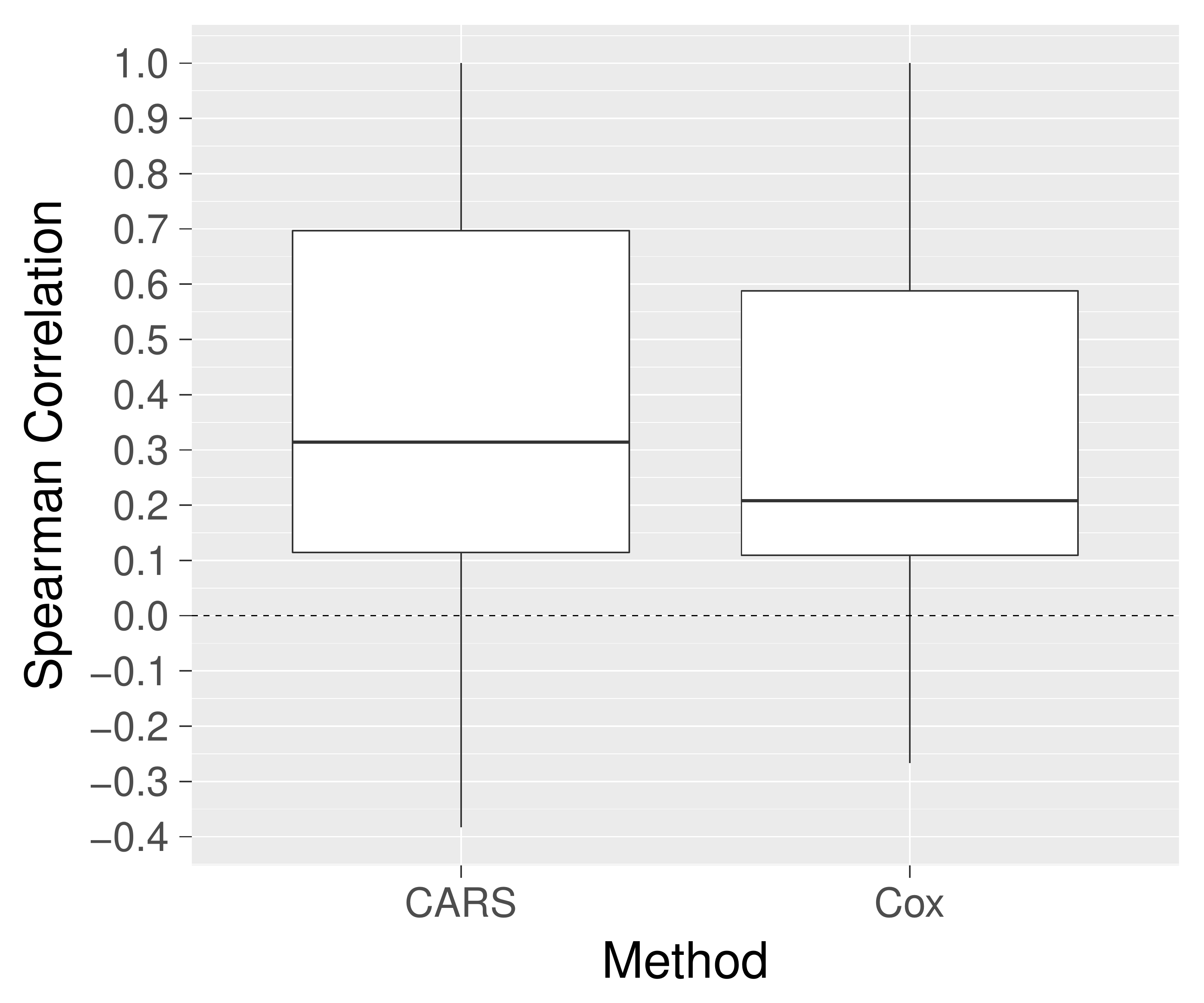}
	\caption{Simulation study -- scenario with low absolute correlations ($\rho = \pm 0.25$):
	This boxplot visualizes the rank correlations of the estimated and the true covariate orderings, as obtained from variable selection by CARS scores, Cox scores and $L_1$-penalized Cox regression. The censoring rate was equal to $25\%$ and low absolute covariate correlations $\rho=\pm 0.25$. Each boxplot shows the results of $24300$ simulation runs (3 explained variance ratios x 3 signal to noise ratios x 3 sample sizes x 3 number of covariates x 300 repetitions).}
	\label{figCor}
\end{figure}

\clearpage 

\subsection{\textbf {Scenario with high absolute correlations and low censoring rate}} 
\label{highCorCase}

 \begin{figure}[!htbp]
 	\centering
   \includegraphics[width=\textwidth]{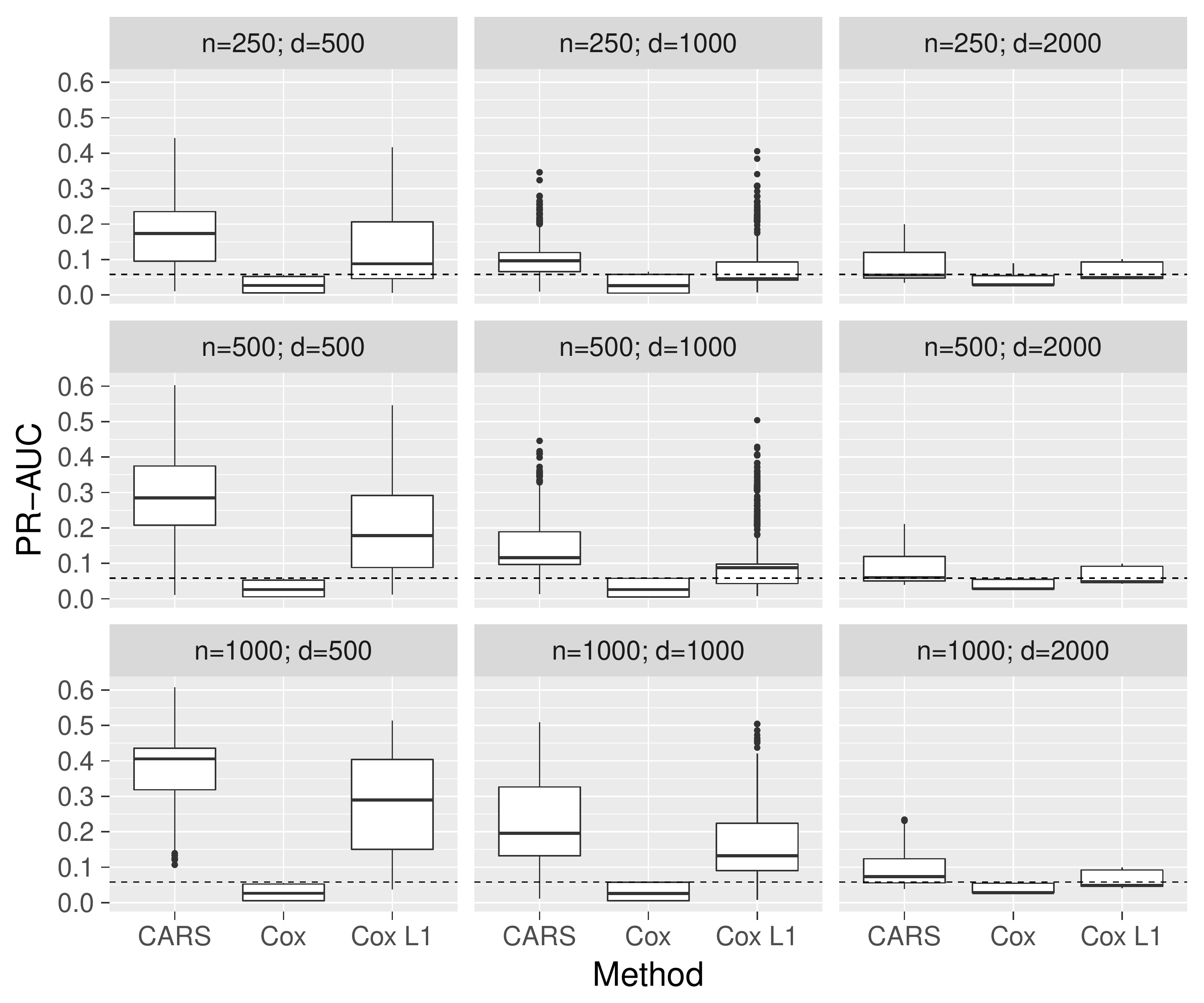}
 	\caption{Simulation: PR-AUC of CARS, Cox and Cox $L_1$ scores stratified by sample size (n) and number of covariates (d) with high absolute covariate correlations $\rho=\pm 0.75$ and censoring rate of $25\%$. Each boxplot shows the results of $2700$ simulation runs (3 explained variance ratios x 3 signal to noise ratios x 300 repetitions).}
 	\label{simCARShighCorPlotNxP}
 \end{figure}
 
  \begin{figure}[!htbp]
  	\centering
    \includegraphics[width=\textwidth]{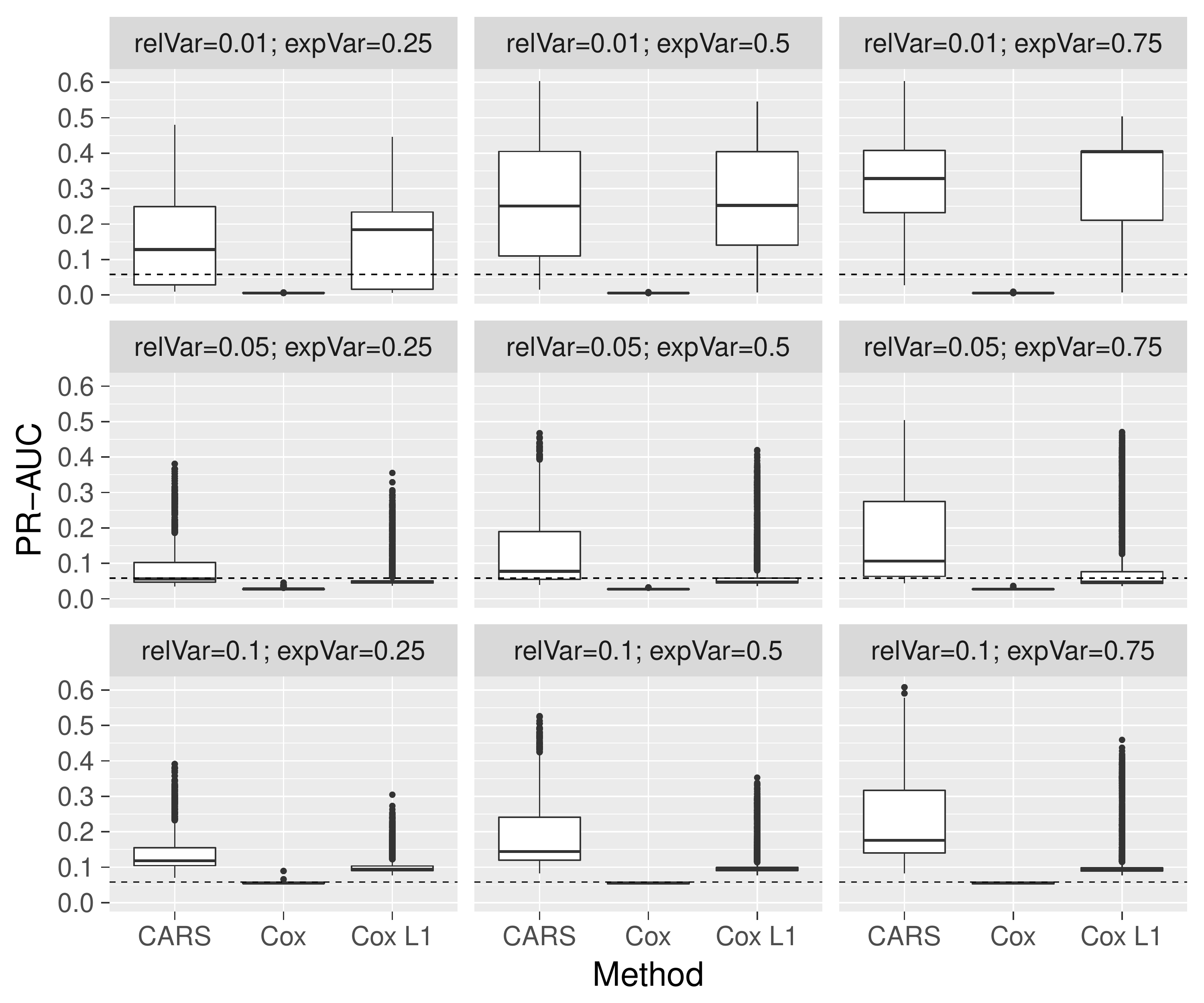}
  	\caption{Simulation: PR-AUC of CARS, Cox and Cox $L_1$ scores stratified by relative number of relevant variables (relVar) and explained variance (expVar) with high absolute covariate correlations $\rho=\pm 0.75$ and censoring rate of $25\%$. Each boxplot shows the results of $2700$ simulation runs (3 sample sizes x 3 number of covariates x 300 repetitions).}
  	\label{simCARShighCorPlotNoVarExpVar}
  \end{figure}

\begin{figure}[!t]
	\centering
  \includegraphics[scale=0.3]{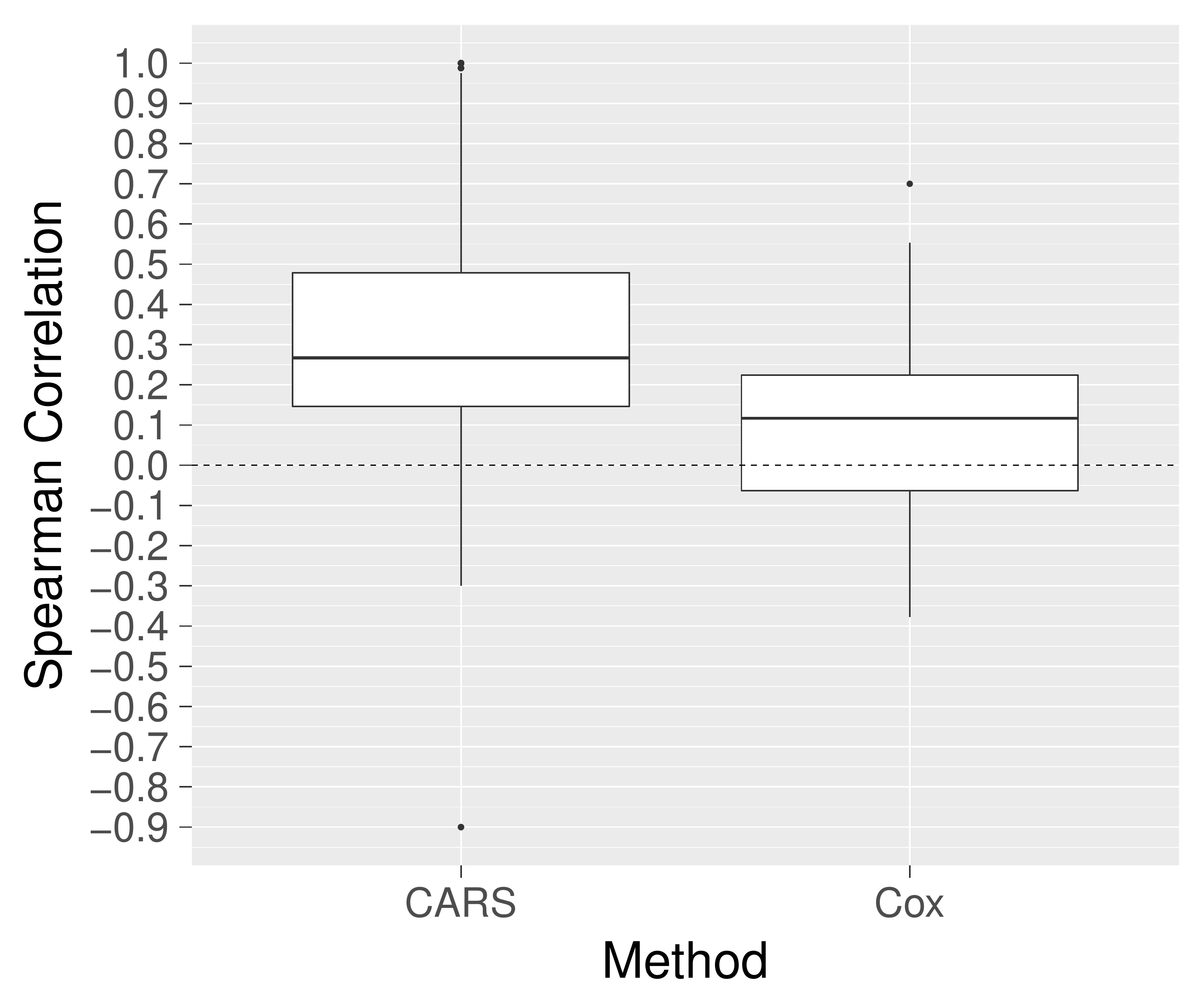}
		\caption{Simulation study -- scenario with high absolute correlations ($\rho = \pm 0.75$):
	The boxplots visualize the rank correlations of the estimated and the true covariate orderings, as obtained from variable selection by CARS scores, Cox scores and $L_1$-penalized Cox regression. The censoring rate was equal to $25\%$ and high absolute covariate correlations $\rho=\pm 0.75$. Each boxplot shows the results of $24300$ simulation runs (3 explained variance ratios x 3 signal to noise ratios x 3 sample sizes x 3 number of covariates x 300 repetitions).}
	\label{figCor-highCor}
\end{figure}

Figure \ref{simCARShighCorPlotNxP} presents the results of the scenario with high absolute correlations (third block of the correlation matrix, $\rho = \pm 0.75$) and with a censoring rate of $25\%$ regarding sample size $n$ and number of covariates $d$. The median of CARS scores was higher than both Cox score approaches for most combinations of $n$ and $d$. However the PR-AUC performance of Cox scores decreased in comparison to the low correlation $\rho=0.25$ scenario. Figure \ref{simCARShighCorPlotNoVarExpVar} displays the results with respect to the number of influential variables  and the explained variances. If $\text{relVar} > 0.05$ CARS scores had again higher median PR-AUC performance than both Cox score approaches. The lack of adjustment for between-covariate correlations degraded performance of Cox scores. This effect propagated to the rank correlation (Figure \ref{figCor-highCor}); in particular, the gap between median rank correlations of CARS and Cox scores had become larger compared to the low correlation scenario (Figure \ref{figCor}). The Figure \ref{simCARShighCorPlotNxP} in the appendix shows a summary of all simulations results.

 \clearpage 
 
\subsection{\textbf {Scenario with high absolute correlations and high censoring rate}}
\label{highCorhighCensCase}

 \begin{figure}[!htbp]
 	\centering
   \includegraphics[width=\textwidth]{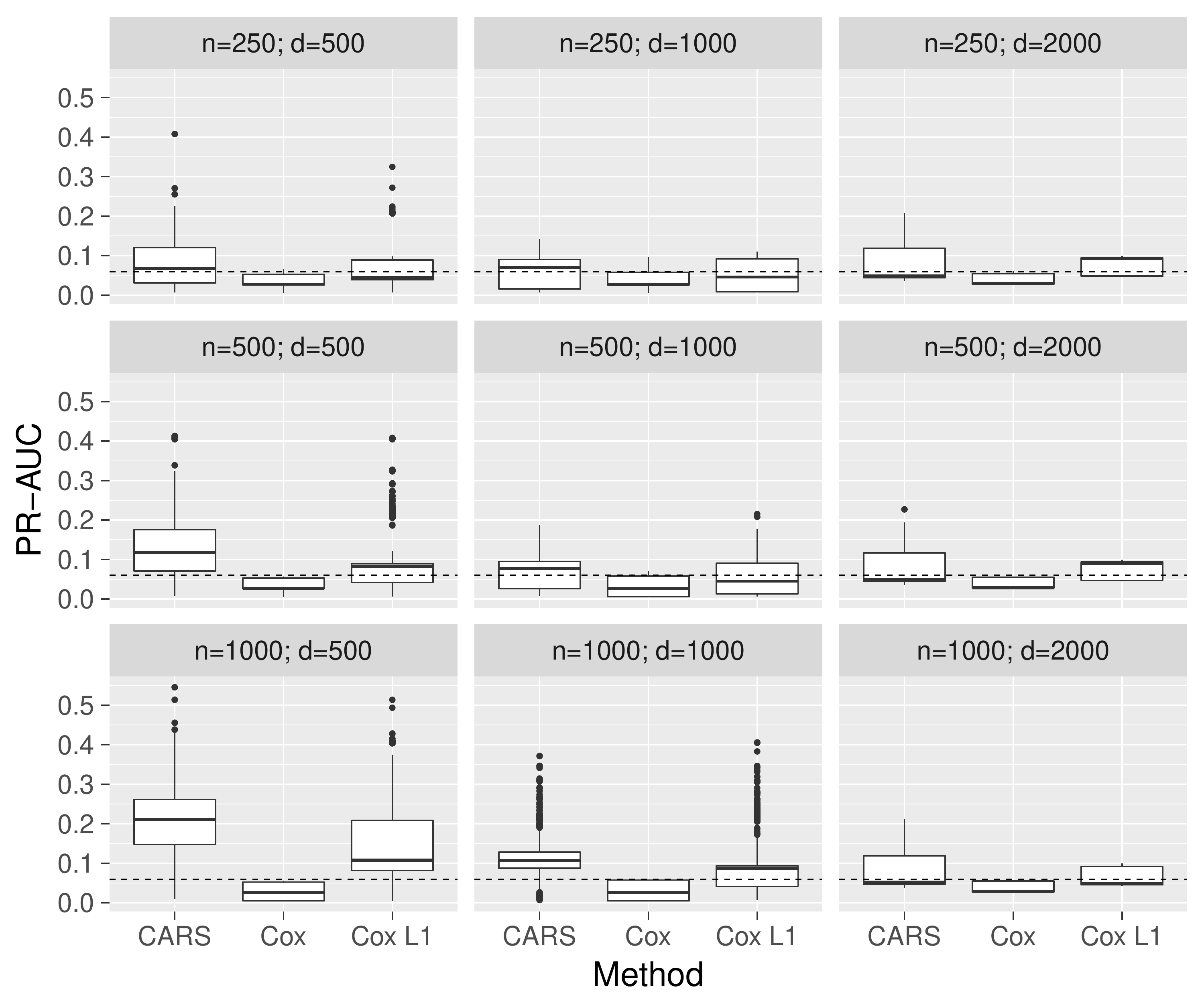}
 	\caption{Simulation: PR-AUC of all methods stratified by sample size (n) and number of covariates (d) with high absolute covariate correlations of $\rho=\pm 0.75$ and censoring rate of $75\%$. Each boxplot shows the results of $2700$ simulation runs (3 explained variance ratios x 3 signal to noise ratios x 300 repetitions).}
 	\label{simCARShighCorHighCensPlotNxP}
 \end{figure}
 
  \begin{figure}[!htbp]
  	\centering
    \includegraphics[width=\textwidth]{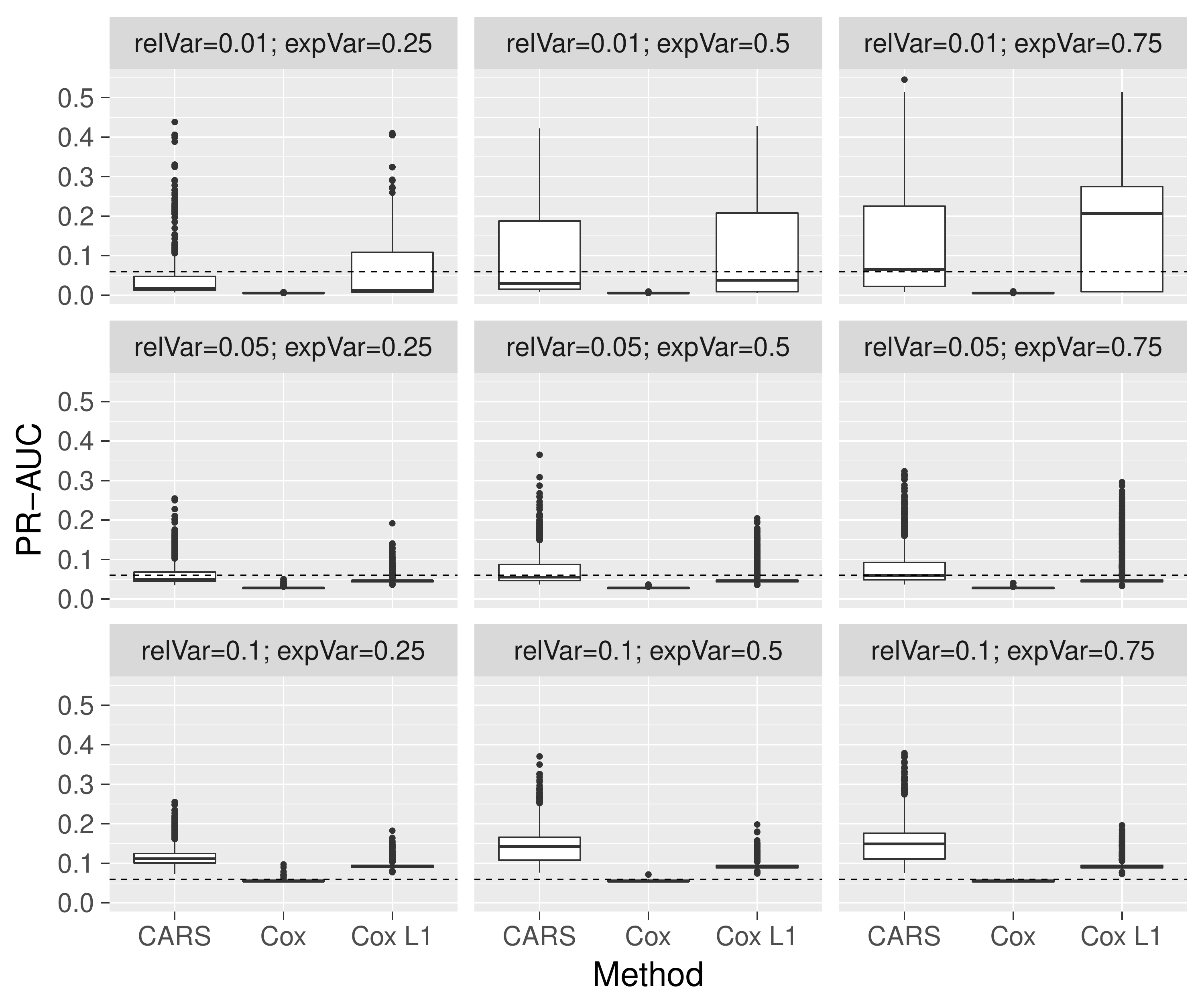}
  	\caption{Simulation: PR-AUC of CARS, Cox and Cox $L_1$ scores stratified by relative number of relevant variables (relVar) and explained variance (expVar) with high absolute covariate correlations of 0.75 and censoring rate of $25\%$. Each boxplot shows the results of $2700$ simulation runs (3 sample sizes x 3 number of covariates x 300 repetitions).}
  	\label{simCARShighCorHighCensPlotNoVarExpVar}
  \end{figure}

\begin{figure}[!htbp]
	\centering
  \includegraphics[scale=0.3]{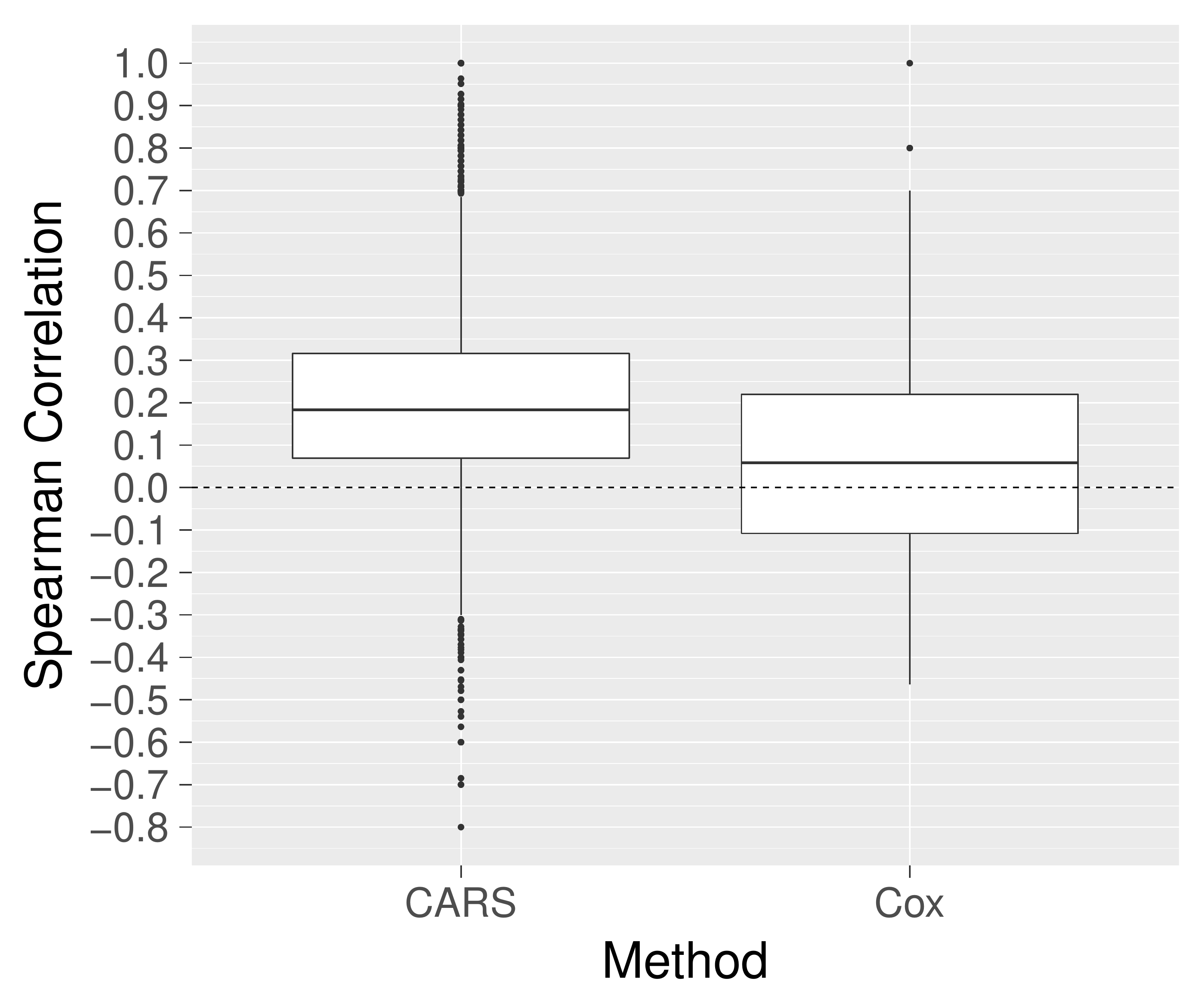}
			\caption{Simulation study -- scenario with high absolute correlations ($\rho = \pm 0.75$) and a high censoring rate of $75\%$:
	The boxplots visualize the rank correlations of the estimated and the true covariate orderings, as obtained from variable selection by CARS and Cox scores. Each boxplot shows the results of $24300$ simulation runs (3 explained variance ratios x 3 signal to noise ratios x 3 sample sizes x 3 number of covariates x 300 repetitions).}
	\label{figCor-highCor-highCens}
\end{figure}

In this section the results of the scenario with high absolute correlations (third block of the correlation matrix, $\rho = \pm 0.75$) and with a high censoring rate of $75\%$ are analyzed. This case is particularly challenging, as approximately $75\%$ of the IPC weights become zero, implying that CARS scores were essentially computed from only $25\%$ of the observations. Consider, for example, the cases $n<=d$ in Figure \ref{simCARShighCorHighCensPlotNxP}, where the PR-AUC performance of all methods was nearly random. If sample size increased above the number of covariates $n>d$ the CARS score median PR-AUC performance grew higher than the Cox approaches compared to the scenarios $n<=d$. Increasing the number of influential covariates $\text{relVaR}=0.01$ to $\text{relVar}=0.1$ yielded better CARS score PR-AUC performance. Especially in cases $\text{relVar}=0.1$ and explained variance $\text{expVar}>=0.5$ CARS scores achieved better median results than Cox approaches. Regarding rank correlations both CARS and Cox approaches behaved similiar as in the previous scenario with high correlations of $\rho=\pm 0.75$ and low censoring rate of $25\%$ (Figure \ref{figCor-highCor-highCens}). The overall summaries with different correlation and censoring structrues, averaged over all design parameters, are given in the Appendix \ref{simCARShighCorHighCens}. In the case with low censoring rates of $25\%$ CARS scores performed better than Cox approaches. High censoring rates of $75\%$ and high absolute covariate correlations of $\rho=\pm 0.75$ CARS scores and Cox scores with $L_1$ penalty were competitive, but without penalty Cox scores were below the PR-AUC of a random classifier. 

\clearpage 

\subsection{\textbf {Runtime in the low correlation, low censoring scenario}} \label{runTime}

\begin{figure}[!htbp]
	\centering
  \includegraphics[width=\textwidth]{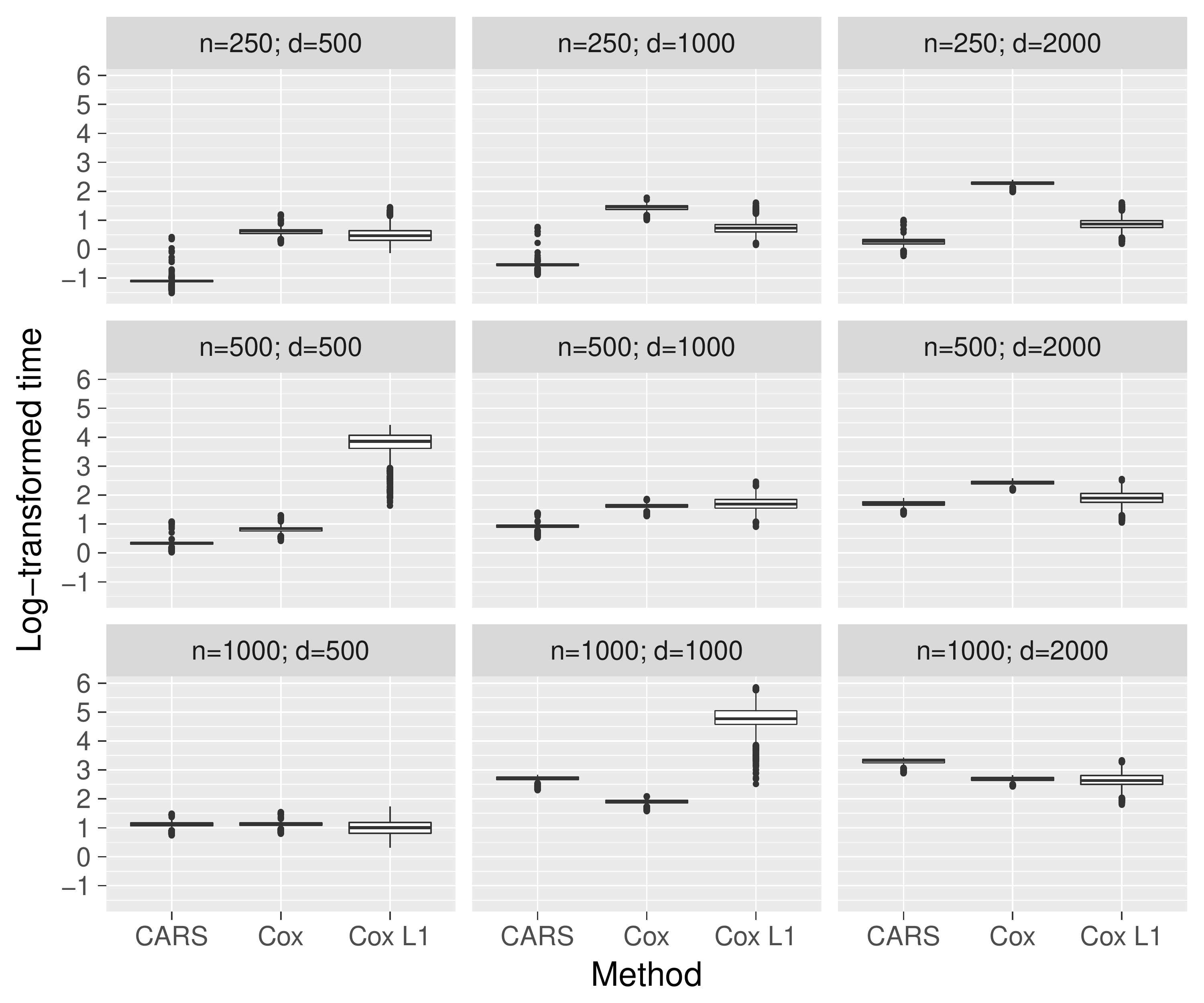}
	\caption{Computational efficiency: Log-transformed time of the three methods CARS, Cox scores and Cox $L_1$ variable selection. The original time was measured in seconds without parallelization. Each boxplot shows the results of $2700$ simulation runs (3 explained variance ratios x 3 signal to noise ratios x 300 repetitions).}
	\label{compTime}
\end{figure}

Besides the predictive performance the computational efficiency of the statistical methods is relevant in the analysis of high-dimensional genomic data. Therefore we additionally measured the computation of CARS, Cox and Cox $L_1$ scores without threshold models in the baseline scenario with low covariate correlations $\rho=\pm 0.25$ and low censoring rate $0.25$. All run times were recorded without parallelization on the same computer with a processor Intel(R) Core(TM) i7-7700 CPU @ 4.20 Ghz and 16 GB RAM in \textit{R} statistical software. CARS scores were computed by using \textit{R} package \textit{carSurv}. Cox scores were calculated by \textit{R} packages \textit{survival} and \textit{glmnet}. Figure \ref{compTime} shows that CARS scores are on average faster to compute than Cox scores with or without $L_1$ penalty for scenarios $n<1000$. Especially in high dimensional context with low number of observations the magnitude between the run times of CARS compared to Cox approaches was large. In scenario $n=1000, d=500$ CARS and Cox approaches yielded comparable runtimes and in the cases $n=1000, d>500$ Cox approaches performed faster.

\clearpage

\subsection{\textbf {Application to the Swedish Watchful Waiting Cohort}} \label{application}

To investigate the properties of the proposed screening method in a real-world setting, we applied the CARS score to the Swedish Watchful Waiting Cohort data \citep{Sboner2010}. The data consists of $281$ patients and $6157$ variables. Beside the clinical covariates (such as patient age, Gleason score and year of diagnosis) an array of $6100$ gene expression profiles (6K DASL) was designed by using four complementary DNA (cDNA)-mediated annealing, selection, ligation, and extension (DASL) assay panels (DAPs) \citep{swedishWatchful_GenePipeline, swedishWatchful_GenePipeline2}. Further details of this procedure are available at GeneExpression Omnibus (GEO: http://www.ncbi.nlm.nih.gov/geo/) with platform accession number \textit{GPL5474}. The data is also available at the GEO website with accession number \textit{GSE16560}. \\

The study population included men who died from prostate cancer during follow up or survived at least 10 years after their diagnosis. The sample size was further restricted to men with high-density tumor regions and who did not receive any type of androgen deprivation. The event of interest was death of prostate cancer; $26.69\%$ of the patients were censored. The median observed time was 100 months (range $\left[6, 274 \right]$ months). The median age was 74 years (range $\left[51, 91 \right]$ years), the median Gleason score was seven (range $\left[2, 10 \right]$), and $58.72\%$ of patients had a lethal diagnosis. The $2.5\%$ and $97.5\%$ quantiles of the Pearson correlations between the gene expressions were $\left[ -0.2634, 0.2865 \right]$ and the maximum absolute correlation was $0.9861$. Therefore most genes were similar to the low correlation used in the simulation design (Section \ref{simulation}), in which CARS scores yielded favorable PR-AUC results. We applied CARS scores to screen for genes that influenced time to death of patients and evaluated their performance in comparison to Cox scores. \\

As the true effects of the genetic markers were unknown, it was not possible to analyze CARS and Cox scores by using the PR-AUC and rank correlation techniques considered in the previous subsections. Instead, we evaluated the scores by comparing their ten times repeated ten-fold-cross-validated predictive performance. The latter was measured by the time-dependent PR-AUC \citep{2016arXiv160604172Y}, which is an extension of PR-AUC to censored data by applying inverse probability weighting \citep{van2012unified}. The time-dependent PR-AUC can be interpreted as average positive predictive value. In addition, we computed a time-independent summary performance measure by weighting and integrating PR-AUC over time ("time-integrated PR-AUC", see Equation \ref{form:kaplanMeierEst} in Appendix \ref{proofOverview}). In each of the $10 \times 10$ training folds CARS and Cox scores were estimated. Each set of risk score values was split into influential and non-influential genetic markers with a predefined q-value cut-off threshold $\alpha_1 \in \left\lbrace 0.01, 0.05, 0.1\right\rbrace$. The cut-off threshold was compared to the q-values given by the method of \cite{fndrNullModel} described in Section \ref{sec2}. For the Cox scores we used the same threshold procedure as CARS scores. All genetic markers with lower q-values than the specified threshold were selected and incorporated into a multivariable Cox regression model. Afterwards, the selected genetic markers were incorporated in a multivariable Cox regression model that also included a clinical baseline formula with the variables age, Gleason score and extremity diagnosis (patient group lethal or indolent) \citep{Sboner2010}. The performance of a random classifier corresponds to the time integrated event rate, which was calculated as the time-dependent event rate $P(T < t_0)$ averaged over all available time points $t_0$ within one fold. The average of the time integrated event rates was computed over $10\times 10$ cross validation folds. \\

\begin{figure}[!htbp]
	\centering
  \includegraphics[width=\textwidth]{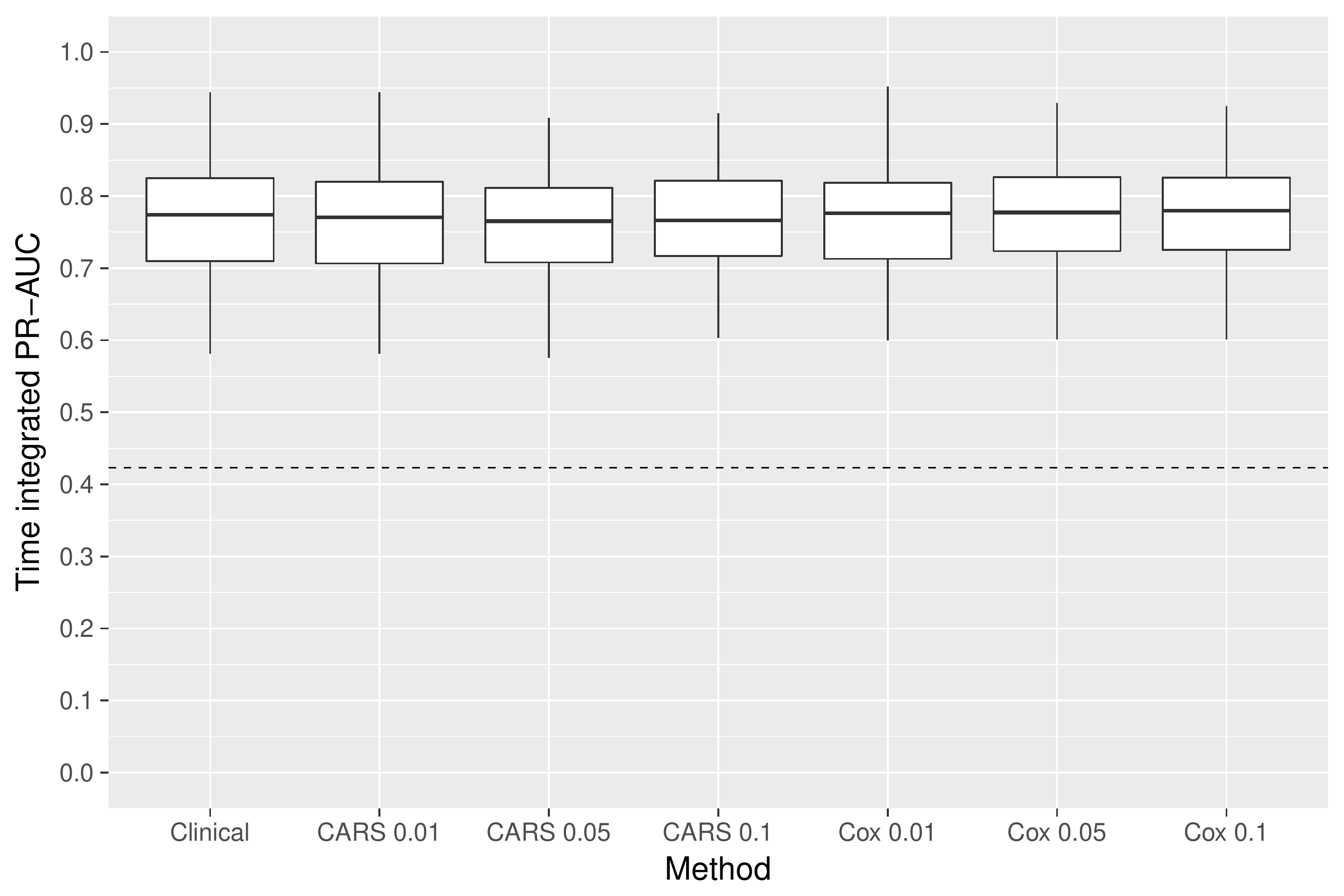}
	\caption{Analysis of the Swedish Watchful Waiting Cohort: The boxplots show the integrated PR-AUCs of Cox models, as obtained from ten times repeated ten-fold cross-validation. In addition to the clinical baseline fomula by \cite{Sboner2010}, the models contained genetic markers that were selected by CARS and Cox scores. The values $\left\lbrace 0.01, 0.05, 0.1\right\rbrace$ represent different q-values cut-off thresholds. The dashed line denotes the average time integrated event rate and corresponds to the performance of a random classifier.}
	\label{swedishWatchful}
\end{figure}

Time integrated PR-AUCs for each fold are shown in Figure \ref{swedishWatchful}. All methods had higher integrated PR-AUC than a random classifier across all cross validation folds. CARS score genetic marker selection resulted in similiar preditive performance compared to genetic marker selection by Cox scores. Both approaches were fairly robust against the choice of the threshold $\alpha_1$. According to Figure \ref{swedishWatchful}, there appears to be no predictive benefit when genetic markers are added to the clinical baseline formula, with CARS scores and Cox scores producing consistent results. This agrees with the findings in the original publication by \cite{Sboner2010}. \\

The complete data analysis with univariate CARS score screening resulted in 0, 3, and 10 identified genetic markers at the q-value thresholds $\alpha_1=\left\lbrace 0.01, 0.05, 0.1 \right\rbrace$, respectively (see appendix Table \ref{tabCompDatAnalysisProst}). Genetic marker selection by Cox scores yielded 1, 1 and 2 genetic markers at the same thresholds. Some of the selected genes by CARS scores with $\alpha_1=0.1$ match previous results from the literature: According to the NCBI database \citep{NCBI}, the  BIRC5 baculoviral IAP repeat containing 5 is an inhibitor of apoptosis and found in most tumor cells. The gene BMX non-receptor tyrosine kinase regulates differentiation and tumorigenicity of several types of cancer cells, and another gene (MLLT11, transcription factor 7 cofactor) was expressed in several leukemic cell lines. The complete list of identified genes with CARS scores is available in the appendix (Table \ref{tabCompDatAnalysisProst}).

\subsection{\textbf {Application to breast cancer microarray data}} \label{appBreast}

In our second real-world example we applied CARS and Cox scores to an invasive breast cancer data set collected by \cite{doi:10.1001/jama.2011.593, Itoh2014}. Merging both available microarray gene expression data sets in the NCBI database \citep[GEO accession numbers \textit{GSE25055}, \textit{GSE25065} and series \textit{GSE25066}]{databaseNCBI} resulted in 502 observations and 22338 variables. These can be partioned into 55 clinical variables, metadata variables and 22283 gene expression markers. The data was collected using \textit{GPL96} [HG-U133A] Affymetrix Human Genome U133A Arrays. The outcome was the time to distant relapse-free survival before surgery (median = 2.716 years, range = $\left[0, 7.439 \right]$ years. 21.91\% of the patients had a relapse within the study duration. The $2.5\%$ and $97.5\%$ quantiles of the Pearson correlations between the gene expressions were $\left[ -0.2467,  0.2725 \right]$ and the maximum absolute correlation was $0.9986$. \\

Analogous to the previous subsection, we used ten times repeated ten-fold cross-validation to analyze predictive performance. The clinical baseline model included the covariates age, tumor stage and an indicator of estrogen receptor (ER) positiveness. The genetic markers were selected by either CARS or Cox scores with different q-value thresholds $\alpha_1=\left\lbrace 0.01, 0.05, 0.1 \right\rbrace$. The significant genetic markers were added to the clinical covariates, and a Cox regression model was fitted. Due to the large number of covariates, Cox regression was regularized with an $L_1$ penalty. The regularization parameter was tuned by internal 10-fold cross-validation as implemented in the \textit{R} package \textit{glmnet}. \\

\begin{figure}[!t]
	\centering
  \includegraphics[width=\textwidth]{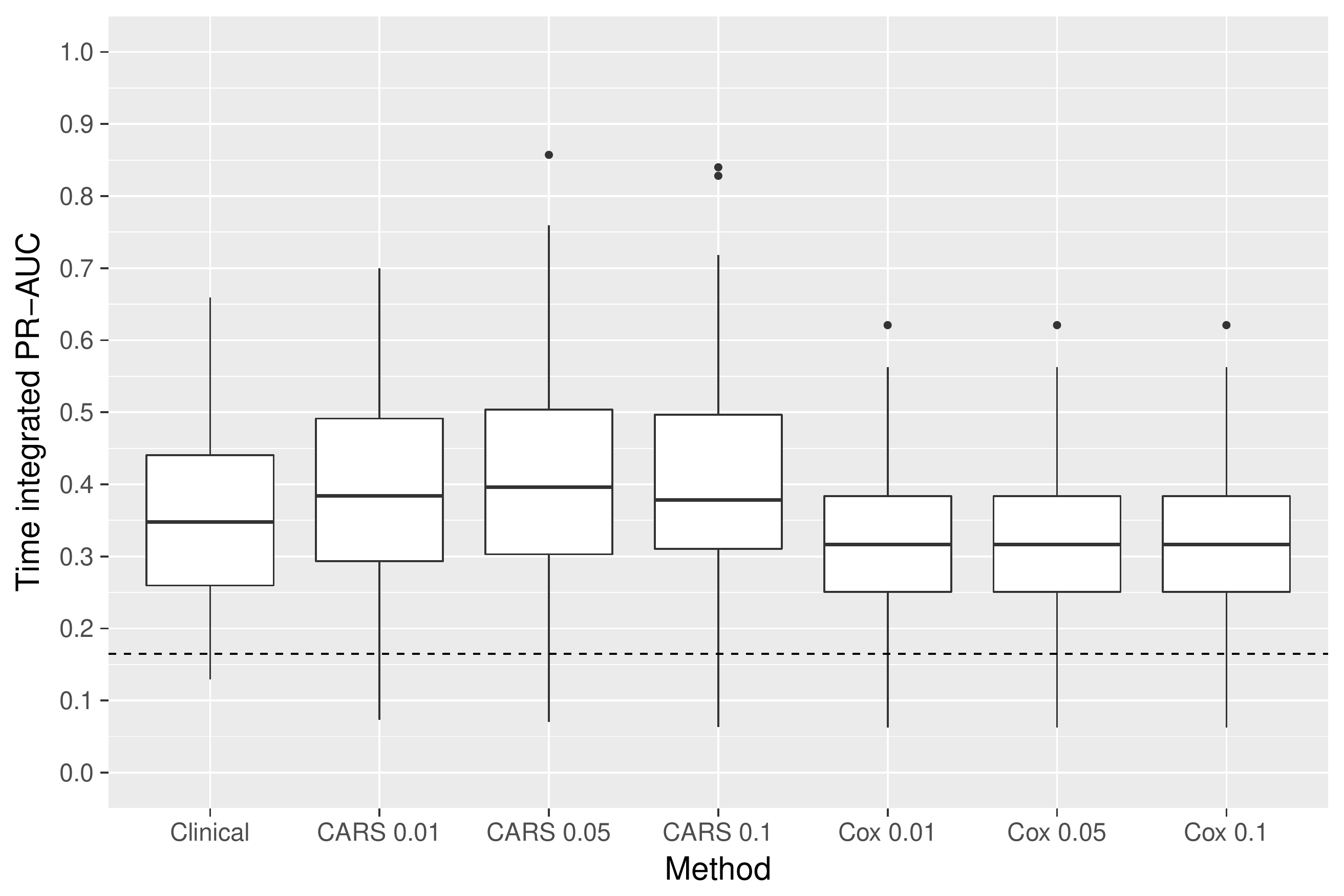}
		\caption{Analysis of the breast cancer data by \cite{doi:10.1001/jama.2011.593}: The boxplots show the integrated PR-AUCs of $L_1$-penalized Cox models, as obtained from ten-times repeated ten-fold cross-validation. In addition to the clinical baseline fomula, the models contained genetic markers that were selected by CARS and Cox scores. The average time integrated event rate is displayed as a dashed line, which corresponds to the performance of a random classifier.
	}
	\label{breastCancer}
\end{figure}

The predictive performance of Cox regression based on CARS and Cox scores is shown in Figure \ref{breastCancer}. It is seen that CARS scores performed better than Cox scores for all levels of $\alpha_1$. For example, when using $\alpha_1=0.01$ as significance threshold, $22$ out of the $22283$ genetic markers were selected by the CARS-based procedure. Genetic marker selection based on Cox scores identified zero genetic markers at $\alpha_1=0.01$ and failed to include influential genetic markers, which degraded predictive performance. In contrast to the data Swedish Watchful Waiting Cohort, there were notable improvements in predictive performance when the genetic markers were added to the clinical model. All identified genes are presented in the appendix (Table \ref{tabCompDatAnalysisBreast1}). \\

In order to annotate the $22$ genes indicated by the CARS score as highly associated with survival at a q-value level $\alpha_1 < 0.01$ we conducted a gene set enrichment analysis based on gene ontology (GO) \citep{Consortium01082001} terms as implemented in the Bioconductor \citep{bioconductor} package \textit{topGO} \citep{Alexa2006}. The GO framework provides a set of structured vocabularies for specific biological domains that can be used to describe gene products in any organism. We computed Fisher's test for enrichment of molecular function and report in supplementary Table \ref{tabTopGO} the $32$ GO terms that were enriched at p-value significance level $\alpha_2 < 0.05$. Among the five GO terms that had attained $\alpha_2 < 0.01$ in the Fisher enrichment test, we detected both protein-glycine ligase activity and protein-glycine ligase activity inhibition. Glycine metabolism has been associated with cancer cell proliferation \citep{Amelio2014}, and glycine uptake and catabolism can promote tumourigenesis and malignancy \citep{Jain2012}. The third enriched GO term was Ras guanyl-nucleotide exchange factor activity. Guanyl-nucleotide exchange factors are proteins that activate GTPases, which are enzymes binding and hydrolizing guanosine triphosphate. Ras is one of the key oncogenes; altough Ras mutations are comparatively rare in breast cancer, the RasGAP (Ras GTPase Activating Proteins) gene RASAL2 functions as tumour suppressor \citep{Laughlin2013}. Furthermore we found enrichment evidence for sodium bicarbonate symporter activity, which enables the transfer of a solute or solutes from one side of a membrane to the other and has a central roles in pH regulation. Solid tumour exhibit different pH profiles compared to normal tissues, which points at a metabolic shift towards acid-producing pathways, reflecting both oncogenic signalling and the development of hypoxia \citep{Gorbatenko2014}. The sodium bicarbonate cotransporter NBCn1 is the predominant mechanism of acid extrusion in primary breast carcinomas compared to normal tissues \citep{Boedtkjer2013} affecting intracellular pH levels. Finally we detected evidence of estrogen 16-alpha-hydroxylase activity, which is one of the earliest reported biomarkers for breast cancer \citep{Bradlow1986}.

%% file: 4-End.tex
\section{\textbf {Summary and discussion}} \label{sec4}

With high-dimensional omics data becoming more readily available in medical research, fast and efficient screening methods are needed for statistical model building and prediction. In this paper, we developed a framework for the selection of genetic markers in time-to-event models. This framework helps to improve biomarker discovery especially in high-dimensional settings with a large number of candidate variables. The proposed CARS score, which evaluates the associations between the de-correlated marker values and the time-to-event outcome, is estimated consistently by combining a set of IPC-weighted variance-covariance estimates. As shown in Section \ref{sec2}, estimates can be computed efficiently even when the number of candidate markers is large. Based on the rankings of the CARS score estimates, genetic markers can be selected for inclusion in a multivariable time-to-event model, where selection errors can be controlled by the adaptive false discovery rate density approach of \cite{fndrNullModel}. \\

In the numerical experiments presented in Section \ref{sec3}, CARS scores showed promising results with regard to the identification of influential marker variables. In particular, screening based on CARS scores outperformed tradional screening methods based on Cox scores in most of the analyzed scenarios. The proposed methodology resulted in increased PR-AUC values, and also in higher correlations between the rankings of the estimated and the true marker effects. With regard to predictive performance, the difference between CARS and Cox scores became largest when marker correlations were high. In these situations, the de-correlation of the markers -- which is the key feature of CARS scores -- had a particularly strong effect on the predictive performance of the multivariable models. Conversely, Cox-based screening -- which ignores the correlations between markers -- could not discriminate between noise and influential variables in low and high covariate correlation settings, thereby degrading predictive performance. Since IPC-weighted estimators tend to be have a high variance when censoring rates are high, we also evaluated the proposed estimators in scenarios with censoring rates as high as $75\%$. Even in these extreme cases, CARS-based screening did not result in a systematically worse performance than Cox-based screening. Furthermore the CARS scores are computationally efficient in high dimensional settings due to exploiting advanced matrix decompositions. \\

CARS scores are based on the theoretical framework of parametric accelerated failure time models. Future research could investigate how to extend this framework to semiparametric and/or nonlinear regression and different censoring mechanisms. 

\section{\textbf {Software}} \label{sec5}

All methods were implemented in \textit{R} \citep{Rsoftware} and published as add-on package \textit{carSurv} (Version 1.0.0), which is available from CRAN. Other packages used in this article include \textit{survival} (Version 2.41-3) \citep{survival-package}, \textit{fdrtool} (Version 1.2.15) \citep{fdrtoolPackage}, \textit{survAUC} (Version 1.0-5) \citep{survAUC}, \textit{ggplot2} (Version 2.2.1) \citep{ggplot2}, \textit{mvnfast} \citep{mvnfast}, \textit{PRROC} (Version 1.3) \citep{PRROC} and \textit{glmnet} (Version 2.0-13) \citep{glmnetPub}. 

\section*{\textbf {Acknowledgements}} \label{Ack}

Financial support from Deutsche Forschungsgemeinschaft (Project SCHM 2966/1-2) is gratefully acknowledged. Verena Zuber is supported by the Wellcome Trust and the Royal Society (Grant Number 204623/Z/16/Z) and the UK Medical Research Council (Grant Number MC\_UU\_00002/7). The authors thank Peter Welchowski for proof reading the manuscript.

\appendix

\section{\textbf {Proof of consistency of CAR survival scores}} \label{proofOverview}

This section proofs that the the CAR survival score $\hat{\btheta}$ is consistent for $\btheta$. CARS scores and their components are defined in the Equations \ref{CARSdef} to \ref{CARSdefLast}. The proof is partitioned into four parts: Second the consistency of the weighted sample variance $s_{Y;w}^2$ of the response is evaluated in the next Section \ref{proofWeightedVar}. Third the consistency of the weighted response sample covariance $s_{X_j,Y;w}$ is analysed (Section \ref{proofWeightedCoVar}). In the last part the consistency of $\hat{\btheta}$ is derived in Section \ref{proofCARS} by combining all previous parts together.

\begin{align} \label{CARSdef}
& \hat{\btheta} = \bR_{\bX}^{-1/2} \bR_{\bX, Y} \\
& \bR_{\bX, Y} = \left(
\frac{s_{X_j,Y;w}}
{\sqrt{s_{X_j}^2} \sqrt{s_{Y;w}^2}}
\right)_{j = 1, \dots , d} \\
& s_{X_j,Y;w} = \frac{1}{n} \sum_{i=1}^n w_{i} (x_{ij} - \bar{x}_j) \left( \log (\tilde{t}_{i} ) - \bar{y}_w \right) \\
& s_{X_j}^2 = \frac{1}{n-1}\sum_{i=1}^n \left( x_{i,j} - \bar{x}_j \right)^2 \\
& s_{Y;w}^2 = \frac{1}{n}\sum_{i=1}^n w_{i} \left( \log (\tilde{t} _{i} ) - \bar{y}_w \right)^2 \\
& \bar{y}_w = \frac{1}{n} \sum_{i=1}^{n} w_i \log (\tilde{t}_{i}) \label{CARSdefLast}
\end{align}

\subsection{\textbf{Consistency of IPC weighted mean}} \label{proofWeightedMean}

To show the consistency of $\bar{y}_w$ it is sufficient to embed this estimator into the framework of unbiased estimation equations \citep{huberEstEq}. A weighted version of the unbiased estimation equations is used to account for censoring, which was developed in the context of nonresponse sample survey theory \citep{surveyTheory, weightedRegSurvey}. In this context the unbiased estimating equation for parameter $\Theta$ is given by

\begin{align} \label{unbEstEqWeight}
& \frac{1}{n} \sum_{i=1}^{n} \psi(\tilde{Y}_i, w_i , \Theta) = 0 \\
& E \left( \psi(\tilde{Y}_i, w_i , \Theta) \right) = 0 \ \forall i=1, \ldots, n \label{expectZero} \\
& w_i = \frac{I(C_i \geq T_i)}{\hat{G}_n(\log(\tilde{T}_i))} \\
& \hat{G}_n(y) = \prod_{j : \log(\tilde{t}_i) \leq y} \left( 1 - \frac{e_j} {r_j} \right) \label{form:kaplanMeierEst}
\end{align}

Under some regularity conditions given in \cite{huberEstEq} (e. g. measureability, continuity and uniqueness of solutions) the estimator of $\hat{\Theta}$ is consistent for $\Theta$ if Equation \ref{expectZero} holds. The weights $w_i$ are inverse probability censoring adjustments based on the ideas of the Horvitz Thompson estimator \citep{horvitzThompson}. Each observation $i=1, \ldots, n$ of the data contains the observed time response, event indicator and covariates $\left(y_i = \log (\tilde{t}_i), \Delta_i, \boldsymbol{x}_i \right)$. The estimate of the logarithmic censoring survivor function $\hat{G}_n(\log(t))$ corresponds to the Kaplan-Meier product estimator \citep{kaplanMeierEst} with individuals at risk $r_j$ just prior the actual $\log(t)$ and number of observed events $e_j$. The Kaplan-Meier product estimator is consistent and therefore the empirical estimate can be replaced by the true survival function in asymptotic analysis. From now on assume that $G(y)$ is known, $G(y) > \nu > 0$ with a small real number $\nu$, $T_i > 0; C_i > 0$ and independence of survival and censoring times. Next consider the estimation equation for the weighted mean $\bar{y}_w$:

\begin{align} 
& \psi(\tilde{Y}_i, w_i , \mu_{\log(T)}) = w_i \log (\tilde{t}_{i}) -  \mu_{\log(T)} \\
& E\left( w_i \log (\tilde{t}_{i}) \right) = \int\limits_{-\infty}^{\infty} \int\limits_{-\infty}^{\infty} \frac{I\left( \log(C_i) \geq \log(T_i)\right) }{G(\log(\tilde{T}_i))} \log(\tilde{T}_i) f_{\log(T)}\left( \log(T)\right) \nonumber \\ 
& f_{\log(C)}\left( \log(C)\right) \ d\log(T) \ d\log(C) 
\end{align}

The indicator function $I\left( \log(C_i) \geq \log(T_i)\right)$ states that only observed survival times contribute to $\psi(\tilde{Y}_i, w_i , \mu_T)$ and therefore $\tilde{T}_i$ can be replaced by $T_i$. The next steps following Equation \ref{unbiasedMean} show that $\bar{y}_w$ is unbiased for the parameter $\mu_{\log(T)}$. It follows under regularity conditions that $\bar{y}_w$ is a consistent estimator of $\mu_{\log(T)}$:

\begin{align} \label{unbiasedMean} 
& E\left( w_i \log (\tilde{t}_{i}) \right) = \int\limits_{-\infty}^{\infty} \frac{1}{G(\log(T_i))} \log(T_i) f_{\log(T)}\left( \log(T)\right) \nonumber \\ & \int\limits_{\log(t_i)}^{\infty} f_{\log(C)}\left( \log(C)\right) \ d\log(T) \ d\log(C) \\
& = \int\limits_{-\infty}^{\infty} \frac{G\left( \log(T_i) \right) }{G(\log(T_i))} \log(T_i) f_{\log(T)}\left( \log(T)\right) \ d\log(T) \\
& = \mu_{\log(T)}
\end{align}

\subsection{\textbf{Consistency of IPC weighted variance}} \label{proofWeightedVar}

The unbiasedness of $\bar{y}_w$ with respect to $\mu_{\log(T)}$ allows to replace $\bar{y}_w$ in the weighted variance estimator $s_{Y;w}^2$ by the true value $\mu_{\log(T)}$ in asymptotic analysis. The $\psi$ transformation for the weighted variance estimator $s_{Y;w}^2$ is given by

\begin{align}
& \psi\left( \tilde{Y}_i, w_i , \sigma_{\log(T)}^2\right)  = w_i \left( \log (T_{i} ) - \mu_{\log(T)} \right)^2 - \sigma_{\log(T)}^2 \\
& = w_i \log (T_{i})^2 - 2 w_i \log (T_{i}) \mu_{\log(T)} + w_i \mu_{\log(T)}^2 - \sigma_{\log(T)}^2 \label{eq:weightVar:Long}
\end{align}

The expectation of the weights $w_i$ equal one and therefore the two middle terms of Equation \ref{eq:weightVar:Long} cancel each other out:

\begin{align}
& E \left( w_i \right) = \int\limits_{-\infty}^{\infty} \int\limits_{-\infty}^{\infty} \frac{I\left( \log(C_i) \geq \log(T_i)\right) }{G(\log(\tilde{T}_i))} f_{\log(T)}\left( \log(T)\right)  f_{\log(C)}\left( \log(C)\right) \ d\log(T) \ d\log(C) \\
& = \int\limits_{-\infty}^{\infty} \frac{1}{G(\log(T_i))} f_{\log(T)}\left( \log(T)\right) \int\limits_{\log(t_i)}^{\infty}  f_{\log(C)}\left( \log(C)\right) \ d\log(T) \ d\log(C) \\
& = \int\limits_{0}^{\infty} \frac{G\left( \log(T_i) \right) }{G(\log(T_i))} f_{\log(T)}\left( \log(T)\right) \ d\log(T) = 1 \\
& \Rightarrow E \left( w_i \left( \log (T_{i} ) - \mu_{\log(T)} \right)^2 \right) =  E(w_i \log (T_{i})^2) - \mu_{\log(T)}^2
\end{align}

The remaining stochastic term $w_i \log (T_{i})^2$ is further evaluated. Using the same reasoning as in Section \ref{proofWeightedMean} the expectation yields unbiasedness of the estimating equation $\frac{1}{n} \sum_{i=1}^{n} \psi\left( \tilde{Y}_i, w_i , \sigma_{\log(T)}^2\right) $ for $\sigma_{\log(T)}^2$ (see Equations \ref{eq:weightVar:Last} to \ref{eq:weightVar:Last2}). Therefore the estimator $s_{Y;w}^2$ is consistent for $\sigma_{\log(T)}^2$. 

\begin{align} \label{eq:weightVar:Last}
& E(w_i \log (T_{i})^2) = \int\limits_{-\infty}^{\infty} \int\limits_{-\infty}^{\infty} \frac{I\left( \log(C_i) \geq \log(T_i)\right) }{G(\log(\tilde{T}_i))} \log (T_{i})^2 f_{\log(T)}\left( \log(T)\right) \nonumber \\  
& f_{\log(C)}\left( \log(C)\right) \ d\log(T) \ d\log(C) \\
& = \int\limits_{-\infty}^{\infty} \frac{1}{G(\log(T_i))} \log (T_{i})^2 f_{\log(T)}\left( \log(T)\right)  \int\limits_{\log(t_i)}^{\infty} f_{\log(C)}\left( \log(C)\right) \ d\log(T) \ d\log(C) \\
& = \int\limits_{-\infty}^{\infty} \log (T_{i})^2 f_{\log(T)}\left( \log(T)\right) \ d\log(T) \\
& = \sigma_{\log(T)}^2 + \mu_{\log(T)}^2 \\
& \Rightarrow E \left( \psi(\tilde{Y}_i, w_i , \sigma_{\log(T)}^2) \right) = 0 \label{eq:weightVar:Last2}
\end{align}

\subsection{\textbf{Consistency of IPC weighted covariance}} \label{proofWeightedCoVar}

Following the same strategy as in the previous Section \ref{proofWeightedVar} the estimating equation is adapted to the sample covariance estimator $s_{X_j,Y;w}$:

\begin{align}
& \psi\left( \tilde{Y}_i, w_i , \sigma_{X_j, \log(T)}\right)  = w_{i} (x_{ij} - \mu_j) \left( \log (\tilde{t}_{i} ) - \mu_{\log(T)} \right) - \sigma_{X_j, \log(T)} \\
& = \underbrace{w_{i} x_{ij} \log (\tilde{t}_{i} )}_{F} - \underbrace{w_{i} x_{ij} \mu_{\log(T)}}_{G} - \underbrace{w_{i} \mu_j \log (\tilde{t}_{i} )}_{H} + w_{i} \mu_j \mu_{\log(T)} - \sigma_{X_j, \log(T)} \label{covABC}
\end{align}

The order of multiple Riemann integrals can be switched without changing the final result \citep{measureTheory}. The analysis of the expectations $F, G, H$ yields 

\begin{align}
& E(F) = \int\limits_{-\infty}^{\infty} \int\limits_{-\infty}^{\infty} \int\limits_{-\infty}^{\infty} \frac{I\left( \log(C_i) \geq \log(T_i)\right) }{G(\log(\tilde{T}_i))} \log(\tilde{T}_i) X_{i,j} f_{\log(T), X_j}\left( \log(T), X_j \right) \nonumber \\ 
& f_{\log(C)}\left( \log(C)\right) \ d\log(T) \ dX_j \ d\log(C)\\
& = \int\limits_{-\infty}^{\infty} \int\limits_{-\infty}^{\infty} \log(T_i) X_{i,j} f_{\log(T), X_j}\left( \log(T), X_j \right) \ d\log(T) \ dX_j = \sigma_{X_j, \log(T)} + \mu_j \mu_{\log(T)} \\
& E(G) = \mu_{\log(T)} \int\limits_{-\infty}^{\infty} \int\limits_{-\infty}^{\infty} \int\limits_{-\infty}^{\infty} \frac{I\left( \log(C_i) \geq \log(T_i)\right) }{G(\log(\tilde{T}_i))} X_{i,j} f_{X_j, \log(T)}\left( X_j,\log(T) \right) \nonumber \\
& f_{\log(C)}\left( \log(C)\right) \ dX_j \ d\log(T) \ d\log(C) \\
& = \mu_{\log(T)} \int\limits_{-\infty}^{\infty} X_j \int\limits_{-\infty}^{\infty} f_{X_j, \log(T)}\left( X_j, \log(T) \right) \ dX_j \ d\log(T) \\
& = \mu_{\log(T)} \int\limits_{-\infty}^{\infty} X_j f_{X_j}\left( X_j \right) \ dX_j = \mu_j \mu_{\log(T)} \\
& E(H) = \mu_j \int\limits_{-\infty}^{\infty} \int\limits_{-\infty}^{\infty} \frac{I\left( \log(C_i) \geq \log(T_i)\right) }{G(\log(\tilde{T}_i))} \log(\tilde{T}_i) f_{\log(T)}\left( \log(T)\right) \nonumber \\
& f_{\log(C)}\left( \log(C)\right) \ d\log(T) \ d\log(C) \\
& = \mu_j \int\limits_{-\infty}^{\infty} \log(T_i) f_{\log(T)}\left( \log(T)\right) \ d\log(T) = \mu_j \mu_{\log(T)}
\end{align}

Combining the results $F, G, H$ with the other terms of the Equation \ref{covABC} shows the unbiasedness of $\psi\left( \tilde{Y}_i, w_i , \sigma_{X_j, \log(T)}\right)$ for the parameter $\sigma_{X_j, \log(T)}$. The consistency follows from the theory of unbiased estimating equations \citep{huberEstEq}: 

\begin{align}
& E(G) = E(H) \\
& E\left( \psi\left( \tilde{Y}_i, w_i , \sigma_{X_j, \log(T)}\right) \right) = E(F) - E(G) - E(H) + \mu_j \mu_{\log(T)} - \sigma_{X_j, \log(T)} \\
& = 0
\end{align}

\subsection{\textbf{Combination of proof results}} \label{proofCARS}

In this section all previous consistency proofs of weighted mean, weighted variance and weighted covariance are combined together to the CARS score. If $\boldsymbol{P}_{\boldsymbol{X}, Y}$ are the true pairwise correlations between the covariates and the response, it follows that

\begin{align}
& \underset{n \rightarrow \infty}{\lim} \boldsymbol{R}_{\boldsymbol{X},Y} = \underset{n \rightarrow \infty}{\lim} \left( \frac{s_{Y, X_j; w}} { s_{X_j} s_{Y; w} } \right)_{j=1, \ldots, d} \longrightarrow \boldsymbol{P}_{\boldsymbol{X}, Y}
\end{align}

 is consistent too, because product and quotient transformations of three consistent estimators are likewise consistent \citep{elemLargeSample}. Another part of the CARS score is the shrinkage estimator $\boldsymbol{R}_{\text{Shrink}}$ of the correlations between the covariates $\boldsymbol{R}_{\boldsymbol{X}}$ \citep{shrinkCovar}

\begin{align}
& \boldsymbol{R}_{\text{Shrink}} = \lambda \boldsymbol{I}_p + (1 - \lambda) \boldsymbol{R}_{\boldsymbol{X}} \\
& \hat{\lambda} = \frac{\sum_{j \neq k} \widehat{Var} (\hat{r}_{j,k})} {\sum_{j \neq k} \hat{r}_{j,k}^2} \\
& \hat{r}_{j,k} = \dfrac{\frac{1} {n-1} \sum_{i=1}^n \left( x_{i,j} - \bar{x}_j \right) \left( x_{i,k} - \bar{x}_k \right)} {\sqrt{\frac{1} {n-1} \sum_{i=1}^n \left( x_{i,j} - \bar{x}_j \right)^2} \sqrt{\frac{1} {n-1} \sum_{i=1}^n \left( x_{i,k} - \bar{x}_k \right)^2}} 
\end{align}

 The shrinkage parameter $\lambda$ is estimated as well. $\boldsymbol{I}_p \in \mathbb{Z}^{p \times p}$ is the unit matrix. $\hat{r}_{j,k}$ are the sample correlations between the j-th and k-th variables. In the limit the estimator converges to zero:

\begin{align}
& \lim_{n \rightarrow \infty} \hat{\lambda} = \frac{\lim_{n \rightarrow \infty} \sum_{j \neq k} \widehat{Var} (\hat{r}_{j,k})} {\lim_{n \rightarrow \infty} \sum_{j \neq k} \hat{r}_{j,k}^2} \\
& = \frac{\sum_{j \neq k} \lim_{n \rightarrow \infty}\widehat{Var} (\hat{r}_{j,k})} {\sum_{j \neq k} \lim_{n \rightarrow \infty} \hat{r}_{j,k}^2} \\
& = \frac{\sum_{j \neq k} 0} {\sum_{j \neq k} r_{j,k}^2} = 0 \\
& \Rightarrow \lim_{n \rightarrow \infty} \hat{\lambda} = 0
\end{align}

Because $\hat{r}_{j,k}$ is a consistent estimator of $r_{j,k}$, the variance of this estimator $\widehat{Var} (\hat{r}_{j,k})$ must converge to zero. Subsequent the shrinkage parameter converges to zero. If $\lambda=0$, then $\boldsymbol{R}_{\text{Shrink}} = \boldsymbol{R}_{\boldsymbol{X}}$ which is itself a consistent estimator of the covariates correlation matrix $\boldsymbol{P}_{\boldsymbol{X}}$. Further the inverse square root $\bR_{\text{shrink}}^{-1/2}$ of the shrinkage estimator is consistent too, because the transformation function is continuous. Combining all previous results shows the consistency of CARS $\hat{\btheta} = \bR_{\text{shrink}}^{-1/2} \bR_{\bX, Y}$ for $\btheta$, because it is a product of two consistent estimators.

\clearpage

\section{\textbf {Additional results obtained from the simulation study}} \label{simCARS}

\subsection{\textbf {CARS simulation with low absolute covariate correlations}} \label{simCARSlowCor}

\begin{figure}[htbp]
	\centering
  \includegraphics[scale = 0.3]{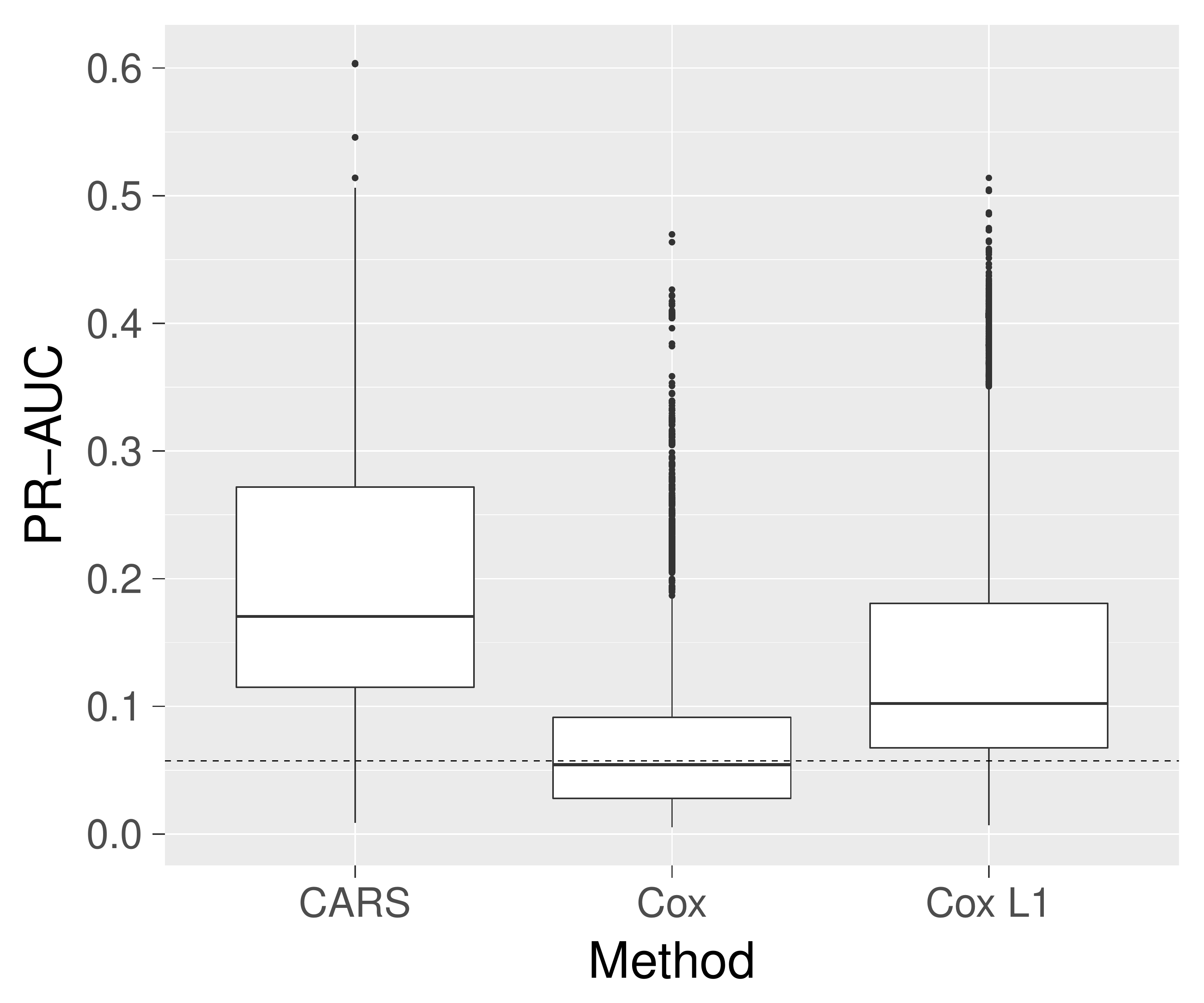}
	\caption{Simulation study -- scenario with low absolute correlations ($\rho = \pm 0.25$): The boxplots visualize the PR-AUC values obtained from variable selection by CARS scores, Cox scores and $L_1$-penalized Cox regression. The censoring rate was equal to $0.25$. The average prevalance of the positive class (computed from all simulations) is displayed by a dashed line. Note that the boxplots contain the PR-AUC values corresponding to all three rates of influential covariates ($1\%$, $5\%$, $10\%$). Each boxplot shows the results of $24300$ simulation runs (3 explained variance ratios x 3 signal to noise ratios x 3 sample sizes x 3 number of covariates x 300 repetitions).}
	\label{figPRAUC}
\end{figure}

\clearpage

\subsection{\textbf {CARS simulation with high absolute covariate correlations}} \label{simCARShighCor}

\begin{figure}[htbp]
	\centering
  \includegraphics[scale=0.3]{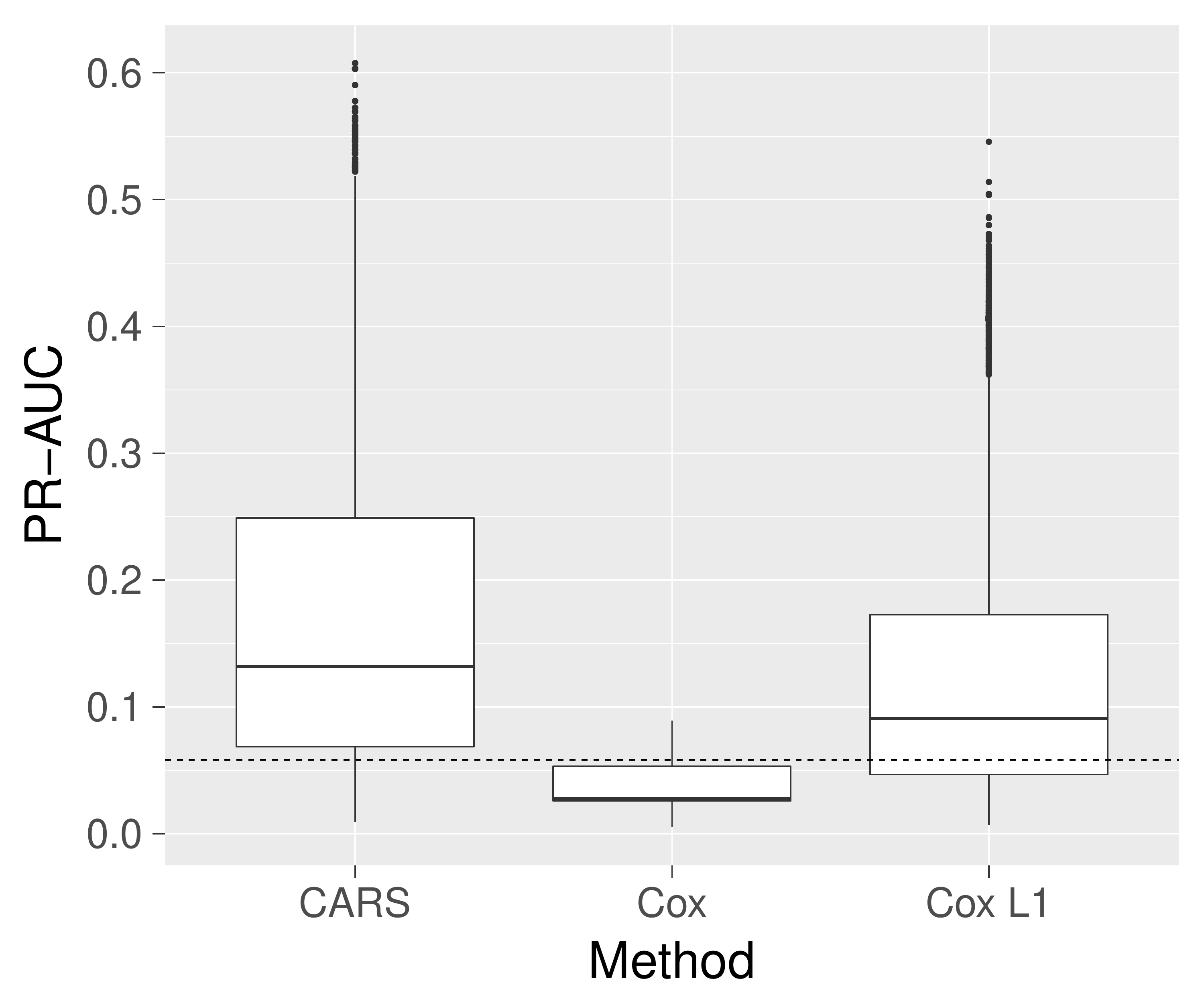}
	\caption{Simulation study -- scenario with high absolute correlations ($\rho = \pm 0.75$): The boxplots visualize the PR-AUC values obtained from variable selection by CARS scores, Cox scores and $L_1$-penalized Cox regression. The censoring rate was equal to $0.25$. The average prevalance of the positive class (computed from all simulations) is displayed by a dashed line. Note that the boxplots contain the PR-AUC values corresponding to all three rates of influential covariates ($1\%$, $5\%$, $10\%$). Each boxplot shows the results of $24300$ simulation runs (3 explained variance ratios x 3 signal to noise ratios x 3 sample sizes x 3 number of covariates x 300 repetitions). Cox scores interquartile range is below PR-AUC of a random classifier.} 
	\label{figPRAUC-highCor}
\end{figure}

\clearpage

\subsection{\textbf {CARS simulation with high absolute covariate correlations and censoring rate 0.75}} \label{simCARShighCorHighCens}

\begin{figure}[htbp]
	\centering
  \includegraphics[scale=0.3]{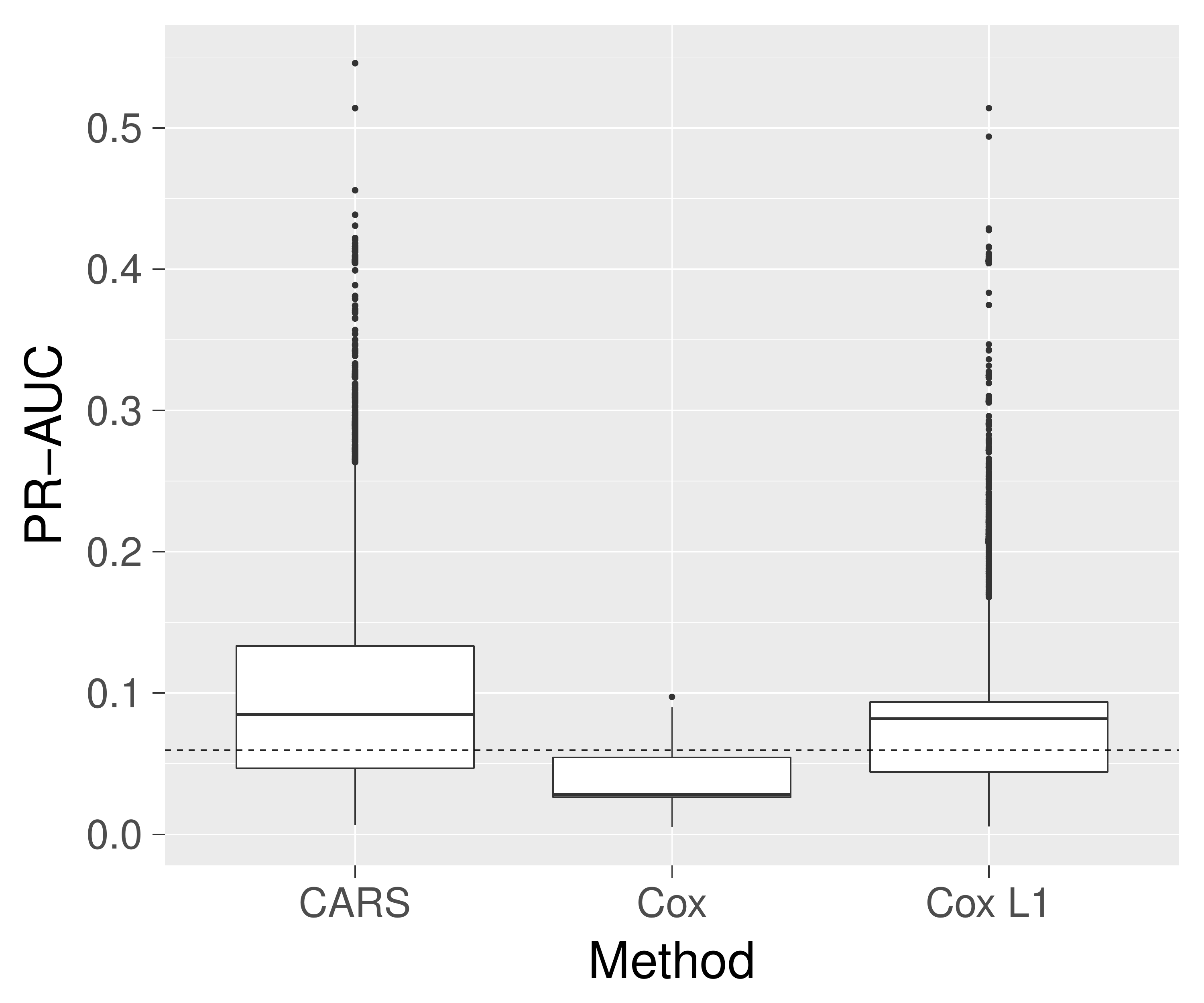}
		\caption{Simulation study -- scenario with high absolute correlations ($\rho = \pm 0.75$) and a high censoring rate of 0.75: The boxplots visualize the PR-AUC values obtained from variable selection by CARS scores, Cox scores and $L_1$-penalized Cox regression. The average prevalance of the positive class (computed from all simulations) is displayed by a dashed line. Note that the boxplots contain the PR-AUC values corresponding to all three rates of influential covariates ($1\%$, $5\%$, $10\%$). Each boxplot shows the results of $24300$ simulation runs (3 explained variance ratios x 3 signal to noise ratios x 3 sample sizes x 3 number of covariates x 300 repetitions). Cox scores interquartile range is below PR-AUC of a random classifier.}
	\label{figPRAUC-highCor-highCens}
\end{figure}

\clearpage

\subsection{\textbf {Construction of correlation matrix}} \label{constrCorr}

In this section give more details for the construction of the correlation matrices used in the simulation (see Section \ref{simulation}). First a preparatory design matrix $\boldsymbol{A}$ is created.  $\boldsymbol{A}$ is partitioned in three blocks of equal size. The correlations between covariates cover a small range in the first block, medium range in the second block and a larger range in the third block. Within each block, half of all correlations are positive and the other half is negative. For example consider the case with 12 covariates: Then each design block would be given by the matrix

\begin{align}
\boldsymbol{A} = \left( 
\begin{matrix}
1 & \xi & \xi & -\xi \\
\xi & 1 & \xi & -\xi \\
\xi & \xi & 1 & -\xi \\
-\xi & -\xi & -\xi & 1 \\
\end{matrix} 
\right)
\end{align}

with $\xi=\left\lbrace 0.25, 0.5, 0.75 \right\rbrace$ in the corresponding blocks. All correlations between variables belonging to different blocks are set to zero. Then the prepared matrix is converted to the nearest possible positive definite matrix $\boldsymbol{B}$, measured by a weighted Frobenius Norm $\left\| \boldsymbol{B} \right\|_W$ of the elementwise differences between the specified and new matrix:

\begin{align}
& \Omega \left( \boldsymbol{B}\right) = \min \left\lbrace \left\| \boldsymbol{A} - \boldsymbol{B} \right\|_W: \boldsymbol{B} \text{ is a correlation matrix} \right\rbrace  \\
& \left\| \boldsymbol{B} \right\|_W = \left\| \boldsymbol{W} \circ \boldsymbol{B} \right\|_F; \ \boldsymbol{W} \text{ symmetric and positive}\\
& \left\| \boldsymbol{B} \right\|_F^2 = \sum_{i,j} b_{i,j}^2
\end{align}

Here $\circ$ denotes the Hadamard product (elementwise matrix multiplication) and $b_{i,j}$ are the elements of matrix $\boldsymbol{B}$. For further details of the algorithm to minimize the deviations $\Omega\left( \boldsymbol{B}\right)$ from $\boldsymbol{A}$ we refere to \cite{nearPD}. A histogram with relative frequencies of the correlation matrix entries, after applying the algorithm, above the diagonal based on $1000$ covariates are shown in the next figure:

\begin{figure}[!htbp]
 	\centering
   \includegraphics[width=\textwidth]{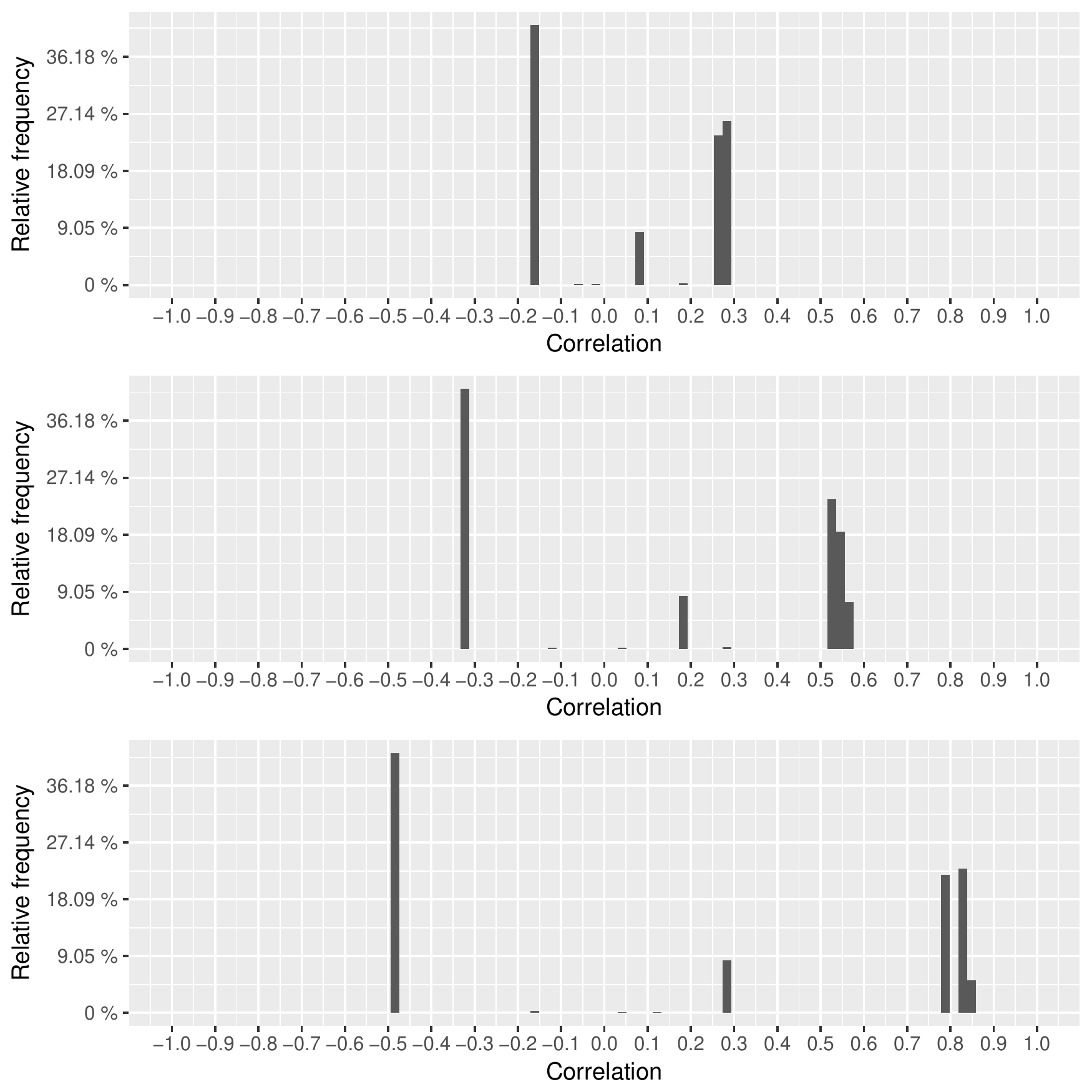}
 	\caption{The top figure shows the histogram (100 equal sized intervals) of the first block of the final correlation matrix with $1000$ covariates. Only correlations above the diagonal are shown. Middle figure shows the histogram of the second block and the lower figure gives the relative frequencies of the third block.}
 	\label{relFreqCorrSim}
\end{figure}

\clearpage

\section{\textbf{Additional material used for the data analysis}}

\subsection{\textbf{CARS scores: Selected variables in data analysis}}

\begin{table}[htbp] \label {tabCompDatAnalysisProst}
	\centering
		\begin{tabular}{|c|r|r|}
			\hline
			Gene symbol & CARS score & Q-value \\
			\hline
			NM\_007244 & 0.1063 & 0.0461 \\
			\hline
			NM\_004912 & -0.1035 & 0.0461 \\
			\hline
			NM\_203281 & -0.1016 & 0.0461 \\
			\hline
			NM\_001012271 & -0.0944 & 0.0814 \\
			\hline
			NM\_021992 & -0.0939 & 0.0835 \\
			\hline
			NM\_006818 & 0.0939 & 0.0836 \\
			\hline
			NM\_005722 & -0.0928 & 0.0877 \\
			\hline
			NM\_003855 & -0.0923 & 0.0895 \\
			\hline
			NM\_001186 & -0.0916 & 0.0918 \\
			\hline
			NM\_006846 & 0.0906 & 0.0965 \\
			\hline
		\end{tabular}
		\caption{Prostate cancer: CARS scores and q-values below $\alpha=0.1$ ordered by absolute score magnitude. Gene symbol refers to the accession number in the NCBI database \citep{databaseNCBI}.}
\end{table}

\begin{table}[htbp] \label {tabCompDatAnalysisBreast1}
	\centering
		\begin{tabular}{|c|r|r|}
			\hline
			Gene Symbol & CARS score & Q-value \\		
			\hline
			GRIN2C & -0.0479   & $1.0560 * 10^{-6}$  \\
			RGP1 & 0.0446   & $6.8075 * 10^{-6}$ \\ 
			TSPAN5 & 0.0446   & $6.8776 * 10^{-6}$ \\ 
			RGS12 & 0.0441 & $7.7700 * 10^{-6}$ \\  
			ITGA2B & 0.0404 & $1.1334 * 10^{-4}$  \\ 
			SURF1 & -0.0402  & $1.3162 * 10^{-4}$ \\ 
			ZC2HC1A & 0.0386 & $3.7071 * 10^{-4}$  \\ 
			MCM9 & 0.0369   & $1.0399 * 10^{-3}$  \\ 
			TRRAP & 0.0368   & $1.0732 * 10^{-3}$  \\ 
			SLC4A5 & 0.0360  & $1.7160 * 10^{-3}$ \\ 
			FEZ2 & -0.0356   & $2.1184 * 10^{-3}$ \\ 
			ARPC4 /// ARPC4-TTLL3 /// TTLL3 & 0.0347 & $3.2360 * 10^{-3}$ \\ 
			KERA & 0.0347  & $3.3192 * 10^{-3}$ \\ 
			GPR98 & 0.0341  & $4.4523 * 10^{-3}$ \\ 
			SEPT6 & 0.0336  & $5.6147 * 10^{-3}$ \\ 
			PRUNE2 & 0.0334 & $6.1200 * 10^{-3}$ \\ 
			PLEKHG3 & 0.0332  & $6.6953 * 10^{-3}$ \\ 
			AW973834$^1$ & 0.0329 & $7.4615 * 10^{-3}$ \\ 
			CYP2C8 & 0.0326  & $8.2900 * 10^{-3}$  \\ 
			CUZD1 & 0.0326  & $8.4730 * 10^{-3}$ \\ 
			CTSF & 0.0324  & $8.8681 * 10^{-3}$   \\ 
			KIAA0485 & 0.0323  & $9.0860 * 10^{-3}$ \\ 
			\hline
		\end{tabular}
		\caption{Breast cancer: CARS scores and q-values below $\alpha=0.01$ ordered by absolute score magnitude. Gene symbol refers to the accession number in the NCBI database \citep{databaseNCBI} or proteins.}
\end{table}

\clearpage

\subsection{\textbf{Summary of gene enrichment analysis}}

\begin{table}[htbp] \label{tabTopGO}
	\centering
		\begin{tabular}{|c|c|r|r|r|r|}
			\hline
			GO.ID &	Term &	Ann &	Signif	& Expect & p-value \\
			\hline
			 GO:0070735	& Protein-glycine ligase activity &	2 &	1 &	0 &	0.0020 \\
			 GO:0070736	& Protein-glycine ligase activity &	2 &	1 &	0 &	0.0020 \\
			 GO:0005088	& Ras guanyl-nucleotide exchange factor & 373 &	3 &	0.35 &	0.0057 \\
			 GO:0008510	& Sodium:bicarbonate symporter activity	& 10 &	1 &	0.01 & 0.0099 \\
			 GO:0101020	& Estrogen 16-alpha-hydroxylase activity & 10 &	1 &	0.01 & 0.0099 \\
			 GO:0005085	& Guanyl-nucleotide exchange factor activity & 470 & 3 & 0.44 & 0.0107 \\
			 GO:0004972	& NMDA glutamate receptor activity & 13	 & 1 &	0.01 & 0.0128 \\
			 GO:0070051	& Fibrinogen binding &	13	& 1	& 0.01	& 0.0128 \\
			 GO:0033695	& Oxidoreductase activity acting on CH & 15	& 1 &	0.01 & 0.0148 \\
			 GO:0034875	& Caffeine oxidase activity	& 15 & 1 & 0.01 & 0.0148 \\
			 GO:0016881	& Acid-amino acid ligase activity & 21 & 1 & 0.02 & 0.0207 \\
			 GO:0005452	& Inorganic anion exchanger activity & 22 & 1 & 0.02 & 0.0217 \\
			 GO:0015106	& Bicarbonate transmembrane transporter activity & 	23 & 1 & 0.02 & 0.0226 \\
			 GO:0016725	& Oxidoreductase activity, acting on CH & 23 & 1 & 0.02 & 0.0226 \\
			 GO:0015301	& Anion:anion antiporter activity & 26 & 1 & 0.02 & 0.0255 \\
			 GO:0004970	& Ionotropic glutamate receptor activity & 29 & 1 & 0.03 & 0.0284 \\
			 GO:0005234	& Extracellular-glutamate-gated ion channel & 30 & 1 & 0.03 & 0.0294 \\
			 GO:0008324	& Cation transmembrane transporter activity & 703 & 3 & 0.66 & 0.0309 \\
			 GO:0004129	& Cytochrome-c oxidase activity & 33 & 1 & 0.03 & 0.0323 \\
			 GO:0015002	& Heme-copper terminal oxidase activity & 33 & 1 & 0.03 & 0.0323 \\
			 GO:0016676 & Oxidoreductase activity & 33 & 1 & 0.03 & 0.0323 \\
			 GO:0008391	& Arachidonic acid monooxygenase activity & 34 & 1 & 0.03 & 0.0333 \\
			 GO:0008392 & Arachidonic acid epoxygenase activity & 34 & 1 & 0.03 & 0.0333 \\
			 GO:0016675 & Oxidoreductase activity & 34 & 1 & 0.03 & 0.0333 \\
			 GO:0070330 & Aromatase activity & 34 & 1 & 0.03 & 0.0333 \\
			 GO:0017112	& Rab guanyl-nucleotide exchange factor activity & 39 & 1 & 0.04 & 0.0381 \\
			 GO:0008066	& Glutamate receptor activity & 42 & 1 & 0.04 & 0.0409 \\
			 GO:0016712	& Oxidoreductase activity & 42 & 1 & 0.04 & 0.0409 \\
			 GO:0022824	& Transmitter-gated ion channel activity & 44 & 1 & 0.04 & 0.0429 \\
			 GO:0022835	& Transmitter-gated channel activity & 44 & 1 & 0.04 & 0.0429 \\
			 GO:0015296	& Anion:cation symporter activity & 46 & 1 & 0.04 & 0.0448 \\
			 GO:0008395	& Steroid hydroxylase activity & 50 & 1 & 0.05 & 0.0486 \\
			\hline
		\end{tabular}
		\caption{Breast cancer: Gene enrichment analyis based on Fisher's test with \textit{R} package \textit{topGO}. The first column gives the gene ontology (GO) \citep{Consortium01082001} identification number and column "Term" gives additional details. The number in column "Ann" shows how many genes are annotated with the GO term. Column "Signif" gives the number of significant genes of the GO term with respect to q-value $0.05$. "Expect" shows the expected number of significant genes under the null hypothesis, e.g. no genes are enriched. The last column "p-value" gives the p-value for Fisher's test for enrichment.}
\end{table}

\clearpage

\subsection{\textbf{CARS scores diagnostic plots}} 

\begin{figure}[!htbp]
	\centering
  \includegraphics[width=\textwidth]{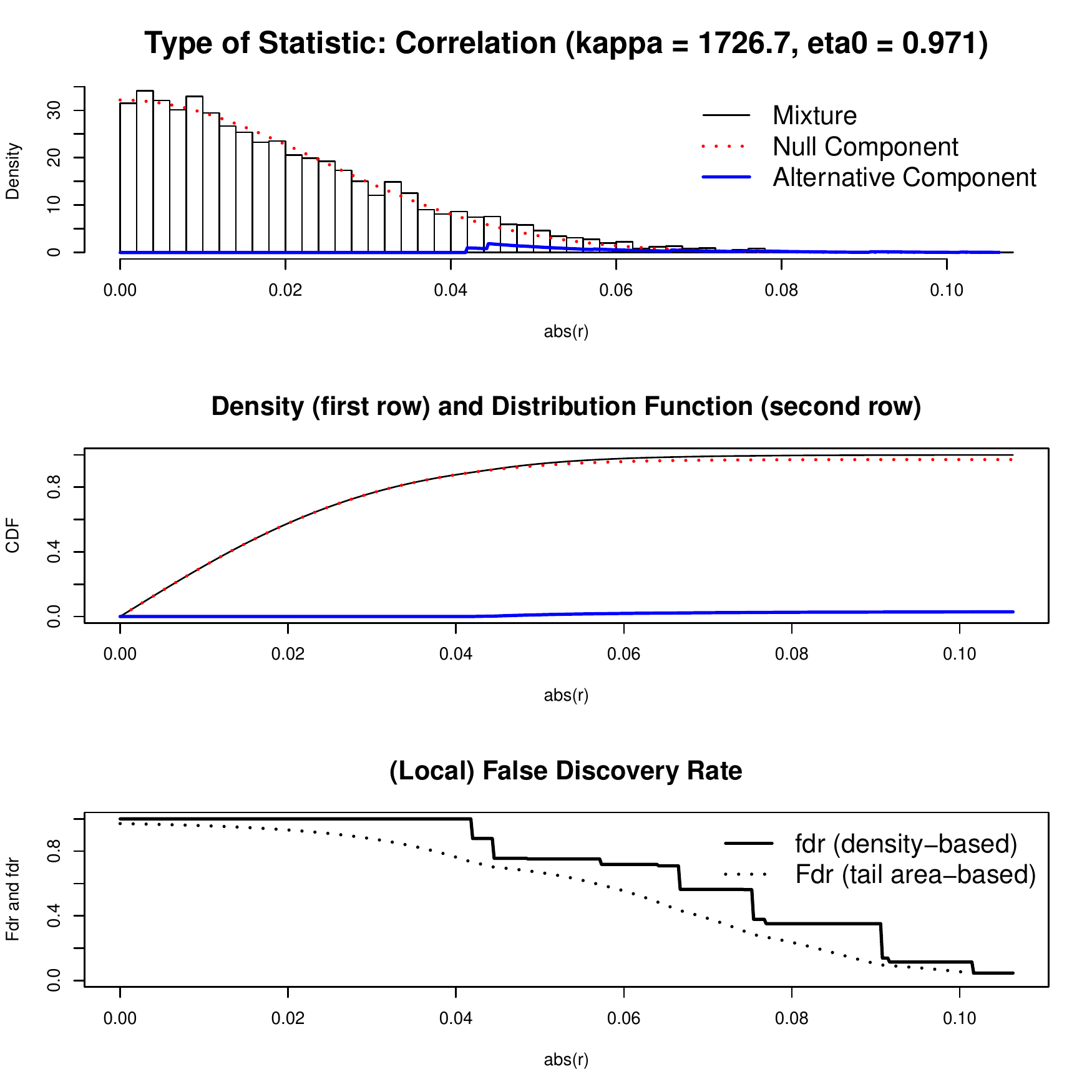}
	\caption{Prostate Cancer model diagnostic plots of CARS scores: Density of empirical null model, cumulative distribution function and local FDR. The first two plots show that the mixture models fits the empirical distribution well.}
	\label{nullModel}
\end{figure}

\begin{figure}[!htbp]
 	\centering
   \includegraphics[width=\textwidth]{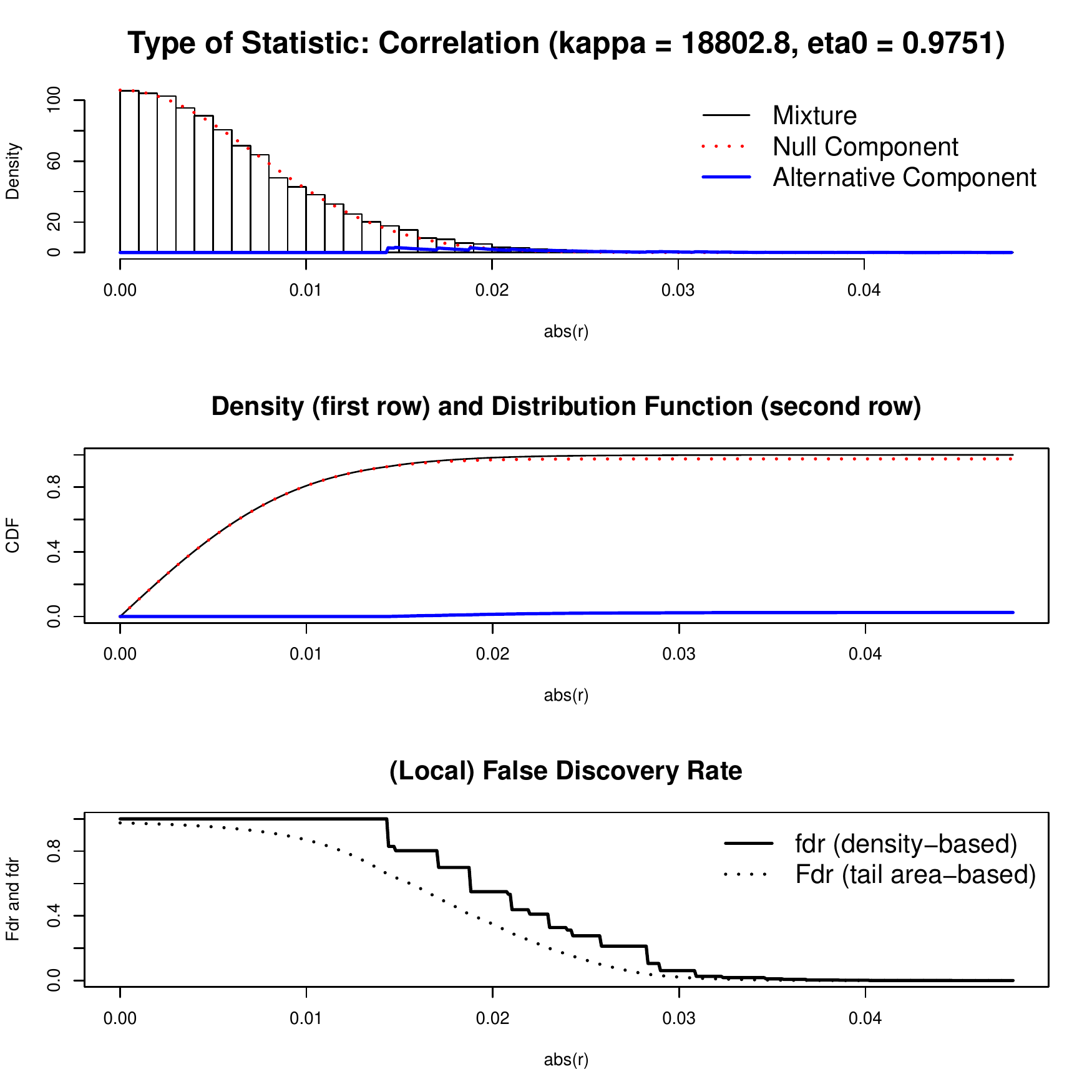}
 	\caption{Breast Cancer model diagnostic plots of CARS scores: Density of empirical null model, cumulative distribution function and local FDR. The first two plots show that the mixture models fits the empirical distribution well.}
 	\label{BreastNullModel}
\end{figure}

\clearpage